\documentclass[aps,prb,reprint,amsmath,amssymb,superscriptaddress,floatfix]{revtex4-2}
\usepackage{graphicx}
\usepackage{bm}
\usepackage{natbib}
\usepackage{url}
\usepackage{textcase}
\usepackage{amssymb}
\usepackage{amsmath}
\usepackage{physics}
\usepackage{appendix}
\usepackage{xcolor}
\usepackage[colorlinks=true,linkcolor=blue,urlcolor=blue,citecolor=blue,anchorcolor=blue]{hyperref}

\begin{document}

\title{
Photoemission Signatures of Photoinduced Carriers and Excitons in One-Dimensional Mott Insulators
}

\author{Taiga Nakamoto}
\affiliation{Department of Physics, The University of Tokyo, Hongo, Tokyo 113-0033, Japan}

\author{Yuta Murakami}
\thanks{These authors supervised this work equally.}
\affiliation{Institute for Materials Research, Tohoku University, Sendai 980-8577, Japan}
\affiliation{RIKEN Center for Emergent Matter Science (CEMS), Wako 351-0198, Japan}

\author{Naoto Tsuji}
\thanks{These authors supervised this work equally.}
\affiliation{Department of Physics, The University of Tokyo, Hongo, Tokyo 113-0033, Japan}
\affiliation{RIKEN Center for Emergent Matter Science (CEMS), Wako 351-0198, Japan}
\affiliation{Trans-Scale Quantum Science Institute, University of Tokyo, Bunkyo-ku, Tokyo 113-8656, Japan}

\date{\today}

\begin{abstract}
We theoretically study photoemission spectra of photodoped one-dimensional Mott insulators that can host an excitonic bound state of a doublon and a holon known as a Mott-Hubbard exciton.
We show that their spectral characteristics differ qualitatively from those of photodoped semiconductors. 
In conventional semiconductors, photoemission spectra are well understood; free charge carriers generate spectral weight near the bottom of the conduction band, while the formation of excitons leads to replica features of the valence band appearing inside the band gap. 
In one-dimensional Mott insulators, on the other hand, strong correlations give rise to fractionalized elementary excitations---spinons, holons, and doublons---which fundamentally modify the photoemission response. 
We find that when photodoped carriers, i.e., doublons and holons, remain unbound, the photoemission spectrum directly reflects the dispersion of spinons, i.e., magnetic elementary excitations. 
In contrast, when a doublon and a holon form a Mott-Hubbard exciton, replica structures of the lower Hubbard band emerge within the Mott gap, carrying contributions from both spinon and holon excitations.
Importantly, the distribution of the in-gap signal depends sensitively on the degree of doublon-holon binding. The origin of these spectral features is clarified through a combination of exact diagonalization and the slave-particle approach. 
These results indicate that photoemission from photoinduced carriers and excitons in strongly correlated electron systems can provide information on magnetic properties and carrier-binding properties.
\end{abstract}
\maketitle

\section{Introduction}
Excitons are bound states of positive and negative charge carriers in solids. They dominate the linear and nonlinear optical properties of insulating materials at subgap energies.
Conventionally, exciton physics has been discussed primarily in the context of semiconductors, where an electron in the conduction band binds to a hole in the valence band due to the attractive Coulomb interaction between them \cite{Frenkel_Transformation_1931,Wannier_Structure_1937}. 
Excitons in semiconductors have attracted considerable attention due to their technological relevance in applications, such as solar cells~\cite{Bernardi_Extraordinary_2013}, photodetectors~\cite{Lopez-Sanchez_Ultrasensitive_2013}, and light-emitting devices~\cite{Ross_Electrically_2014}, as well as their connection to novel phenomena including exciton condensation~\cite{Morita_Observation_2022}.

On the other hand, the concept of excitons is not limited to conventional semiconductors but is important in understanding optical properties of strongly correlated electron systems (SCESs)~\cite{Stephan_Dynamical_1996,gallagher_Excitons_1997,Hanamura_Excitons_2000,Essler_Excitons_2001,Wrobel_Excitons_2002,Tohyama_Resonant_2002,Jeckelmann_Optical_2003,Matsueda_Excitonic_2005,Al-Hassanieh_Excitons_2008,nakamoto_Onedimensional_2025}. 
A salient feature of excitons in such systems is that strong electronic correlations can yield properties fundamentally distinct from those of semiconductor excitons~\cite{Murakami_Photoinduced_2025}. 
In particular, strongly correlated materials often exhibit characteristic magnetic features, which can have profound effects on excitonic behavior. 
For example, in one-dimensional Mott insulators exhibiting spin–charge separation, the dipole moment between bright and dark excitons becomes anomalously large, leading to giant third-order optical responses~\cite{Kishida_Gigantic_2000,Ono_Linear_2004,Ono_Direct_2005,Baykusheva_Quantum_2026}.
Furthermore, in higher-dimensional Mott insulators, excitons can be formed not primarily through the Coulomb interaction, as in conventional semiconductors, but rather through magnetic fluctuations~\cite{Terashige_Doublonholon_2019,Mehio_Hubbard_2023}. 
In addition, recent experiments have reported the existence of unconventional excitons coupled to magnetic excitations in two-dimensional van der Waals magnetic materials such as NiPS$_3$~\cite{Kang_Coherent_2020,Dirnberger_Spincorrelated_2022}.
Understanding excitons in strongly correlated materials, particularly their intimate relationship with magnetic structures, is therefore crucial for uncovering excitonic physics beyond the conventional semiconductor framework and for opening new avenues toward novel optical functionalities.

\begin{figure*}[t]
  \centering
  \includegraphics[width=0.90\textwidth]{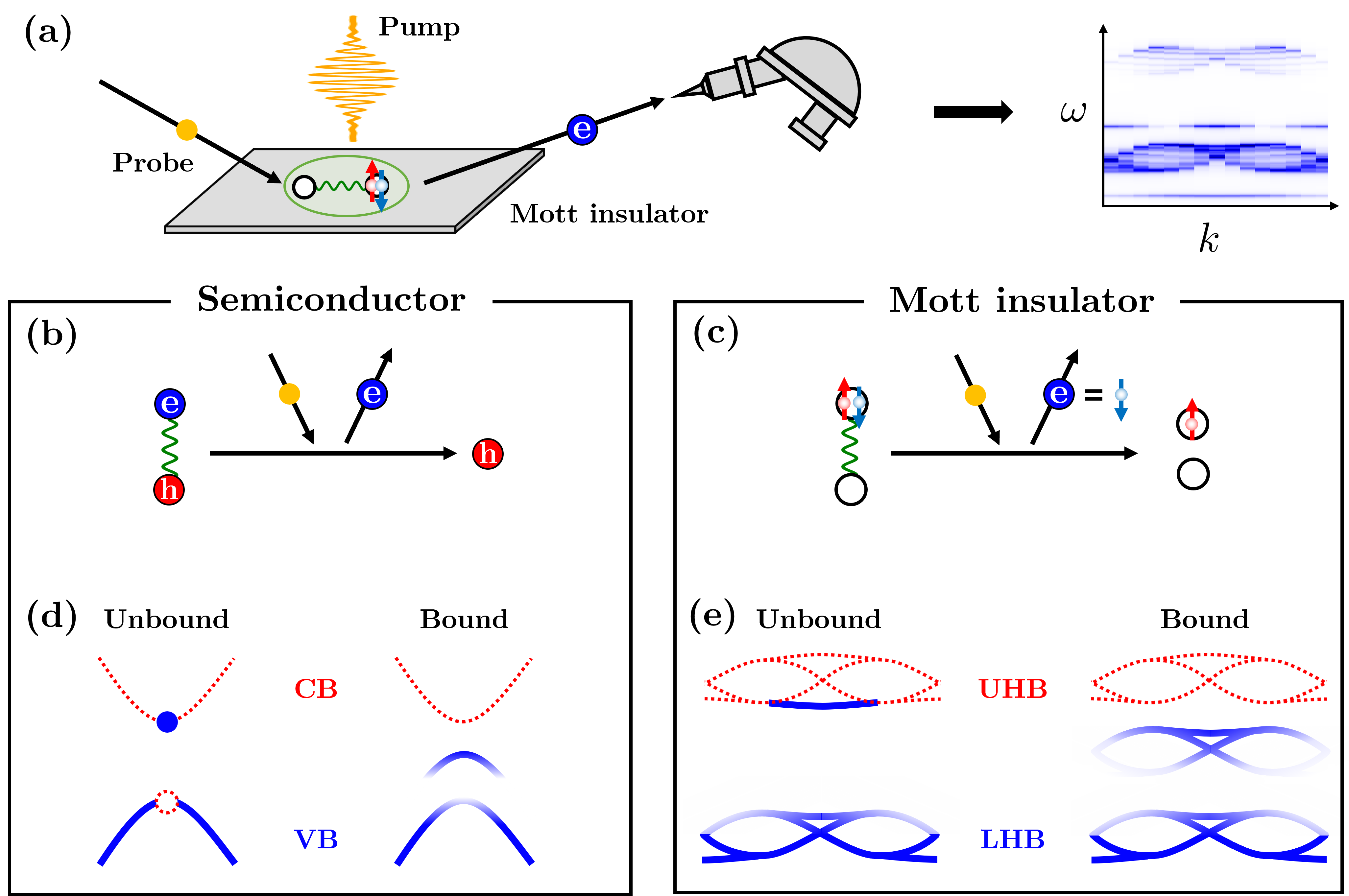}
  \caption{
    (a) Schematic picture of photoemission spectroscopy for photodoped Mott insulators, which are induced by a pump pulse, creating doublons (doubly occupied sites) and holons (empty sites).
    If the interaction between doublons and holons is strong enough, they form bound states called Mott-Hubbard excitons.
    One can measure the momentum-resolved photoemission spectrum by applying a probe pulse, where electrons are emitted from the photodoped Mott insulator.
    (b) Photoemission process for semiconductor excitons, where photons emit electrons and holes remain in the system. 
    (c) Photoemission process for Mott-Hubbard excitons, where photons emit electrons from doublons, and holons and spin excitations (spinons) remain in the system.
    (d) Schematic picture of the photoemission spectrum of photodoped semiconductors.
    In the case that electrons and holes are unbound, their photoemission signal appears at the bottom of the conduction band (CB).
    When electrons and holes are bound, replica structures of the valence band (VB) appear below the CB.
    (e) Schematic picture of the photoemission spectrum of photodoped Mott insulators.
    In the case that doublons and holons are unbound, a dispersive band signal appears just below the upper Hubbard band (UHB).
    When doublons and holons are bound, replica structures of the lower Hubbard band (LHB) appear below the UHB.
    } \label{fig:Concept}
\end{figure*}

Exciton studies in both semiconductors and SCESs have been pursued mainly from an optical perspective. 
In recent years, however, time- and angle-resolved photoemission spectroscopy (trARPES) has emerged as a powerful experimental technique that provides momentum-resolved information on excitons beyond optical probes~\cite{boschini_Timeresolved_2024}. 
Given that an exciton is a composite particle formed by multiple charge carriers, photoemission spectroscopy—which detects electrons emitted from a system by a probe pulse—can be viewed as a solid-state analogue of high-energy collider experiments~\cite{Steinhoff_Exciton_2017}. 
Indeed, in conventional semiconductors, it has been demonstrated that the lineshape of photoemission spectra originating from excitons directly reflects the underlying electron–hole wavefunction~\cite{perfetto_Firstprinciple_2016,Rustagi_Photoemission_2018,Christiansen_Theory_2019,stefanucci_Carriers_2021}. 
Exploiting this relation, recent studies have successfully estimated the spatial extent of excitons in transition-metal dichalcogenides from photoemission measurements~\cite{Man_Experimental_2021, Karni_Structure_2022, Schmitt_Formation_2022, Bange_Ultrafast_2023}. 
In contrast, several theoretical studies have addressed photoemission spectroscopy from excitons in SCESs~\cite{Bittner_Photoenhanced_2020}. 
Nevertheless, a comprehensive understanding remains elusive regarding several key aspects: the similarities and differences relative to semiconductor excitons, the physical origin of spectral features, and the role of underlying magnetic correlations.

As a first step toward a systematic understanding of these issues, we focus here on the one-dimensional extended Hubbard model. 
Compared with higher-dimensional systems, its one-dimensional nature enables the use of systematic numerical and analytical approaches. 
This model therefore provides a minimal theoretical framework for describing excitons in SCESs, and has possible applications to materials such as halogen-bridged nickel compounds and copper oxide chain compounds~\cite{Tokura_RMP}.
By treating photoexcited states as steady states~\cite{Murakami_Photoinduced_2025}, we systematically analyze photoemission spectra for photoexcited Mott insulators (Fig.~\hyperref[fig:Concept]{1(a)}).
One can understand excitons realized in this system, referred to as Mott-Hubbard excitons, as bound states of doublons (doubly occupied sites) and holons (empty sites).
If electrons are emitted from Mott-Hubbard excitons in a photoemission process, holons and spin excitations (spinons) are left in the system (Fig.~\hyperref[fig:Concept]{1(c)}).
In conventional semiconductors, on the other hand, what remain after photoemission from excitons are only holes (Fig.~\hyperref[fig:Concept]{1(b)}).
This difference results in qualitatively different spectral features between Mott-Hubbard excitons and conventional semiconductor excitons.

Our analysis demonstrates that the photoemission signal from unbound doublons and holons already exhibits dispersive in-gap signals below the upper Hubbard band (UHB) (Fig.~\hyperref[fig:Concept]{1(e)}).
These in-gap features reflect the properties of spinon degrees of freedom.
This is in contrast to the case of conventional semiconductors, where the photoemission signal from unbound electrons and holes appears only at the bottom of the conduction band (Fig.~\hyperref[fig:Concept]{1(d)}).
When the Mott-Hubbard excitons are formed, we find an in-gap signal separated from the UHB, whose structure resembles that of the lower Hubbard band (LHB).
Although this signal is analogous to the excitonic band in conventional semiconductors, 
it involves richer physical processes.
Namely, its complex structure reflects the intertwined motion of spinons and holons initiated by the photoemission process.
We discuss the physical origin of the photoemission signal arising from Mott-Hubbard excitons in terms of the exact diagonalization method and the slave-particle method.

This paper is organized as follows. 
In Sec.~\ref{sec:Model}, we introduce the model Hamiltonian and the steady state formulation to describe the photodoped Mott insulator.
In Sec.~\ref{sec:Numerical_Results}, we show the photoemission spectra obtained from the exact diagonalization and point out the characteristic features.
Then, we provide physical interpretations of these features with two theoretical approaches.
In the first approach, we analyze the elementary excitations dynamics after photoemission using the time-dependent Lanczos method (Sec.~\ref{sec:Time-dependent_Lanczos_method}).
In the second approach, we employ the slave-particle method to decompose the photoemission spectrum into charge and spin contributions (Sec.~\ref{sec:Slave-particle_method}).
Finally, we summarize our results in Sec.~\ref{sec:Conclusions}.

\section{Model} \label{sec:Model}
In this work, we focus on the one-dimensional extended Hubbard model,
\begin{align}
    \hat{H} &= -t_{\mathrm{hop}}\sum_{j,\sigma} (\hat{c}^{\dagger}_{j,\sigma}\hat{c}_{j+1,\sigma} + \mathrm{h.c.}) + \hat{H}_U+ \hat{H}_V, \label{eq:1D-Extended_Hubbard_model} \\
    \hat{H}_{\mathrm{U}} &= U\sum_{j}(\hat{n}_{j,\uparrow} - 1/2)(\hat{n}_{j,\downarrow} - 1/2), \\
    \hat{H}_{\mathrm{V}} &= V\sum_{j}(\hat{n}_{j} - 1)(\hat{n}_{j+1} - 1),
\end{align}
where $t_{\mathrm{hop}}$ is the hopping parameter, $U$ ($V$) is the on-site (nearest-neighbor) interaction strength,
$\hat{c}_{j,\sigma}$ ($\hat{c}^{\dagger}_{j,\sigma}$) is the annihilation (creation) operator of electrons with spin $\sigma$ at site $j$, $\hat{n}_{j,\sigma}=\hat{c}^{\dagger}_{j,\sigma}\hat{c}_{j,\sigma}$ is the electron number operator with spin $\sigma$ at site $j$, and $\hat{n}_{j} = \hat{n}_{j,\uparrow} + \hat{n}_{j,\downarrow}$ is the total electron number operator.
This model is often used to describe one-dimensional strongly correlated materials, such as halogen-bridged nickel compounds and copper oxide chain compounds~\cite{Tokura_RMP}.
In this paper, we focus on the half-filled system with $U>2V$ and $U\gg t_{\rm hop}$.
The corresponding ground state is a Mott insulating state, where localized electrons give rise to spin degrees of freedom that exhibit quasi-long-range antiferromagnetic correlations.

When a Mott insulator is photoexcited across the Mott gap, doublon–holon pairs are created. 
If the nearest-neighbor Coulomb interaction $V$ is sufficiently large, a doublon and a holon can bind to form a Mott–Hubbard exciton.
Here, we aim to evaluate the photoemission spectrum in such a photodoped state. 
In principle, the spectrum can be obtained by explicitly simulating the nonequilibrium time evolution after photoexcitation~\cite{Sugimoto_Pumpprobe_2023}. 
However, such simulations are often computationally demanding.
To systematically study the photoemission signal in photodoped systems without resorting to numerically intensive calculations, we adopt a quasi-equilibrium (quasi-steady-state) description of photodoped systems~\cite{Murakami_Photoinduced_2025}.
This approach is motivated by the fact that the recombination time of photocarriers is exponentially long in large-gap Mott insulators~\cite{Strohmaier_Observation_2010,Lenarcic_Ultrafast_2013,Mitrano_PressureDependent_2014,Sensarma_Lifetime_2010,Eckstein_Thermalization_2011,Lenarcic_Charge_2014,Nevola_Timescales_2021}, where initially generated photocarriers can relax within the Hubbard bands via electron scattering and coupling to the environment (e.g., electron–phonon interactions) on timescales shorter than recombination.
The system can therefore enter a long-lived, quasi-steady state, analogous to the situation in photodoped semiconductors~\cite{Rustagi_Photoemission_2018}. 

Previous studies have argued that the quasi-steady state can be approximated by an equilibrium state of an effective Hamiltonian subject to conserved doublon and holon numbers~\cite{Rosch_Metastable_2008,Takahashi_Photoexcited_2002,Takahashi_Photoinduced_2002,Gomi_Photogenerated_2005,Li_eta-paired_2020,Li_Nonequilibrium_2021,Murakami_Exploring_2022,Murakami_Spin_2023,sarkar_Floquet_2024,Ueda_Exotic_2025}.
The effective Hamiltonian can be obtained by the Schrieffer-Wolff transformation \cite{MacDonald_$fractU$_1988} from the original Hamiltonian \eqref{eq:1D-Extended_Hubbard_model} as
\begin{align}
    \hat{H}_{\mathrm{eff}} &= \hat{H}_{\mathrm{kin,d}} + \hat{H}_{\mathrm{kin,h}} + \hat{H}_U + \hat{H}_V + \hat{H}_{\mathrm{s}} + \hat{H}_{\mathrm{\eta}}, \label{eq:Effective_Hamiltonian} \\
    \hat{H}_{\mathrm{kin,d}} &= -t_{\mathrm{hop}}\sum_{j,\sigma}\hat{n}_{j,\bar{\sigma}}(\hat{c}^{\dagger}_{j,\sigma}\hat{c}_{j+1,\sigma}+\mathrm{h.c.})\hat{n}_{j+1,\bar{\sigma}}, \\
    \hat{H}_{\mathrm{kin,h}} &= -t_{\mathrm{hop}}\sum_{j,\sigma}\hat{\bar{n}}_{j,\bar{\sigma}}(\hat{c}^{\dagger}_{j,\sigma}\hat{c}_{j+1,\sigma}+\mathrm{h.c.})\hat{\bar{n}}_{j+1,\bar{\sigma}},  \\
    \hat{H}_{\mathrm{s}} &= J\sum_j\hat{\vb{s}}_j \cdot \hat{\vb{s}}_{j+1}, \\
    \hat{H}_{\mathrm{\eta}} &= -J\sum_j\hat{\vb{\boldsymbol{\eta}}}_j \cdot \hat{\vb{\boldsymbol{\eta}}}_{j+1},
\end{align}
where $\hat{\bar{n}}_{j,\sigma} = (1-\hat{n}_{j,\sigma})$, $\bar{\sigma}$ denotes the spin opposite to $\sigma$, and $J=4t_{\mathrm{hop}}^2/U$ is the spin (doublon-holon) exchange interaction strength.
The spin operator $\hat{\vb{s}}$ is defined as $\hat{\vb{s}} = \sum_{\alpha, \beta}\hat{c}^{\dagger}_{\alpha}\vb{\sigma}_{\alpha,\beta}\hat{c}_{\beta}/2$ with $\vb{\sigma}$ being Pauli matrices.
The $\eta$-spin operator is defined as $\hat{\eta}_i^{+} = (-)^i \hat{c}^{\dagger}_{i,\downarrow}\hat{c}^{\dagger}_{i,\uparrow}$ and $\hat{\eta}_i^{z} = (\hat{n}_i-1)/2$.
$\hat{H}_{\mathrm{kin,d}}$ ($\hat{H}_{\mathrm{kin,h}}$) represents the doublon (holon) kinetic energy term, $\hat{H}_{\mathrm{s}}$ represents the Heisenberg spin interaction term, and $\hat{H}_{\mathrm{\eta}}$ represents the doublon-holon interaction term.

Throughout the paper, we set $t_{\mathrm{hop}}$ as the unit of energy.
We mainly use the parameter $U=20$ in order to clearly see the spectral features related to the Mott-Hubbard exciton in the main text. 
In Appendix~\ref{sec:Spectrum_U10}, we confirm that similar spectral features also appear for $U=10$, a value typical of realistic one-dimensional Mott insulators.  
We also confirm there that a result obtained from the steady-state approach agrees well with that from the direct time evolution~\cite{Sugimoto_Pumpprobe_2023}.
This agreement indicates that our steady-state approach can capture the spectral features after the intraband relaxation process.
We note, however, that our framework cannot describe the transient dynamics during the pump pulse and the subsequent interband relaxation process, and the spectral features may be different during these processes.
The number of spin-up and spin-down electrons $N_{\mathrm{\uparrow}}=N_{\mathrm{\downarrow}}$ is fixed to $L/2$ with $L$ being the system size (i.e., half-filling without spin polarization), and we employ the periodic boundary condition.
For simplicity, we focus on the sector of $N_d=N_h=1$ (i.e. photodoped Mott insulators with one doublon and one holon), and approximate the photodoped state as the lowest energy state of this sector.

We obtain photoemission signals by calculating the single-particle spectral function $A(k,\omega)$ with the effective Hamiltonian $\hat{H}_{\mathrm{eff}}$ as
\begin{align}
    A(k,\omega) &= A^{<}(k,\omega) + A^{>}(k,\omega), \label{eq:def_Spectral_function} \\
    A^{<}(k,\omega) &= \sum_{m,\sigma} \delta(\omega+E_{m0})\abs{\bra{\psi_m}\hat{c}_{k,\sigma}\ket{\psi_0}}^2 , \label{eq:A_lesser} \\
    A^{>}(k,\omega) &= \sum_{m,\sigma} \delta(\omega-E_{m0})\abs{\bra{\psi_m}\hat{c}^{\dagger}_{k,\sigma}\ket{\psi_0}}^2, \label{eq:A_greater}
\end{align}
with $E_{m0}=E_m-E_0$.
Here, $E_m$ and $\ket{\psi_m}$ are the $m$-th eigenenergy and eigenstate of $\hat{H}_{\mathrm{eff}}$, and, in particular, $\ket{\psi_0}$ and $E_0$ represent the photodoped steady state and its energy.
$A^{<}(k,\omega)$ ($A^{>}(k,\omega)$) is the momentum-resolved occupied (unoccupied) spectrum, and $k$ is the crystal momentum.
Note that when the system has the particle-hole symmetry, we have $A^{<}(k,\omega) = A^{>}(\pi-k,-\omega)$.
We are mainly interested in $A^{<}(k,\omega)$, which is directly related to ARPES signals.
We note that experimentally observed ARPES spectra are products of the single-particle spectral functions and photoemission matrix elements.
Since the latter depends on the details of materials and the experimental setup~\cite{boschini_Timeresolved_2024}, we only focus on the single-particle spectral function in this work.

For photodoped Mott insulators, it is convenient to decompose $A^{<}(k,\omega)$ into doublon annihilation (singlon creation) and singlon annihilation (holon creation) processes.
Here "singlon" refers to singly occupied sites.
To this end, we rewrite the electron annihilation operator as
\begin{align}
  \hat{c}_{j,\sigma} &= \hat{c}_{j,\sigma,\mathrm{D\to S}} + \hat{c}_{j,\sigma,\mathrm{S\to H}}, \\
  \hat{c}_{j,\sigma,\mathrm{D\to S}} &= \hat{c}_{j,\sigma}\hat{n}_{j,\bar{\sigma}}, \label{eq:c_DtoS} \\
  \hat{c}_{j,\sigma,\mathrm{S\to H}} &= \hat{c}_{j,\sigma}(1-\hat{n}_{j,\bar{\sigma}}), \label{eq:c_StoH}
\end{align}
where $\hat{c}_{j,\sigma,\mathrm{D\to S}}$ changes doublons into singlons with spin $\bar{\sigma}$, and $\hat{c}_{j,\sigma,\mathrm{S\to H}}$ changes singlons with spin $\sigma$ into holons.
Within the effective-model description, the numbers of doublons and holons are conserved, respectively.
This allows us to decompose $A^{<}(k,\omega)$ as
\begin{align}
  A^{<}(k,\omega) &= A^{<}_{\mathrm{D\to S}}(k,\omega) + A^{<}_{\mathrm{S\to H}}(k,\omega), 
  \label{eq:A_lesser2}\\
  A^{<}_{\mathrm{D\to S}}(k,\omega) &= \sum_{m,\sigma} \delta(\omega+E_{m0})\abs{\bra{\psi_m}\hat{c}_{k,\sigma,\mathrm{D\to S}}\ket{\psi_0}}^2 , \label{eq:A_lesser_DtoS} \\
  A^{<}_{\mathrm{S\to H}}(k,\omega) &= \sum_{m,\sigma} \delta(\omega+E_{m0})\abs{\bra{\psi_m}\hat{c}_{k,\sigma,\mathrm{S\to H}}\ket{\psi_0}}^2 , \label{eq:A_lesser_StoH}
\end{align}
where $\hat{c}_{k,\sigma,\mathrm{D\to S}}$ and $\hat{c}_{k,\sigma,\mathrm{S\to H}}$ are the spatial Fourier transforms of $\hat{c}_{j,\sigma,\mathrm{D\to S}}$ and $\hat{c}_{j,\sigma,\mathrm{S\to H}}$, respectively.
For convenience, we call $A^{<}_{\mathrm{D\to S}}(k,\omega)$ the doublon antihalation spectrum, and call $A^{<}_{\mathrm{S\to H}}(k,\omega)$ the holon creation spectrum.

\begin{figure*}[t!]
  \centering
  \includegraphics[width=0.9\textwidth]{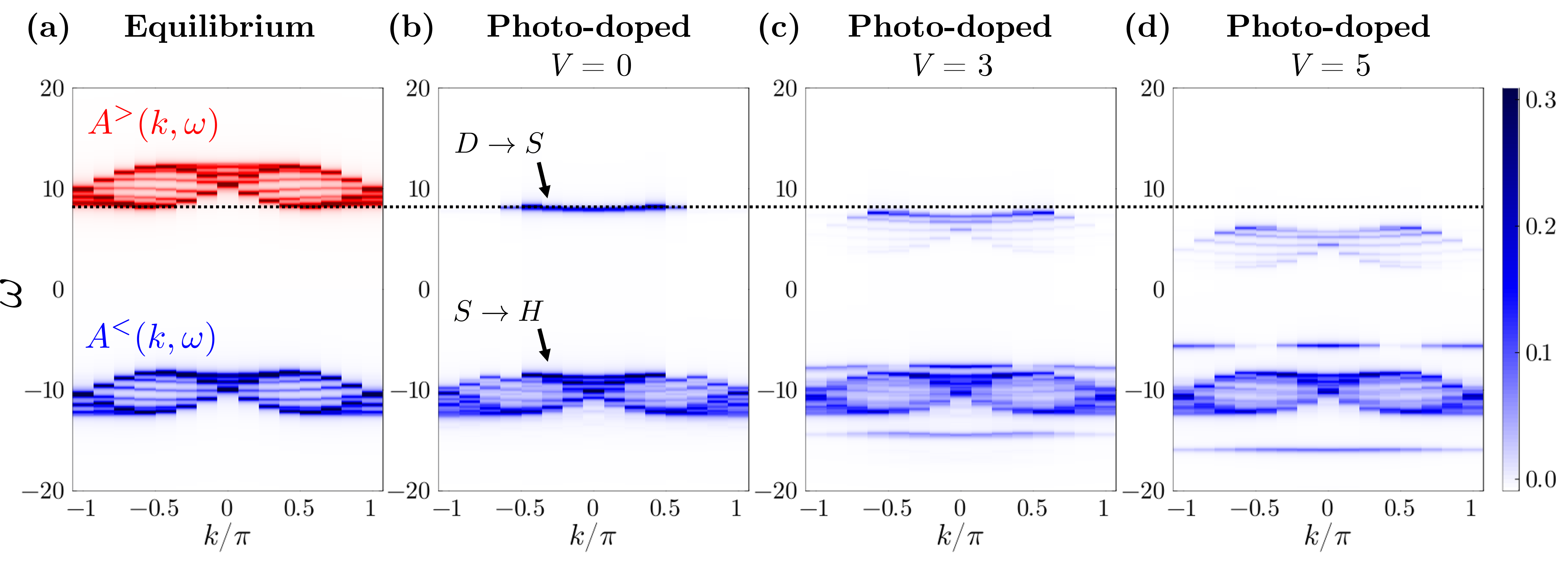}
  \caption{
    (a) Single-particle spectral function $A(k,\omega) = A^{>}(k,\omega)+A^{<}(k,\omega)$ for equilibrium Mott insulators ($N_d=N_h=0$).
    The unoccupied spectrum $A^{>}(k,\omega)$ (red color scale) corresponds to the UHB, while the occupied spectrum $A^{<}(k,\omega)$ (blue) corresponds to the LHB.
    The black dashed line shows the band edge of the UHB.
    (b-d) Occupied spectrum $A^{<}(k,\omega) = A^{<}_{\mathrm{D\to S}}(k,\omega) + A^{<}_{\mathrm{S\to H}}(k,\omega)$ for photodoped Mott insulators ($N_d=N_h=1$) with (b) $V=0$, (c) $V=3$, and (d) $V=5$.
    $A^{<}_{\mathrm{D\to S}}(k,\omega)$ has intensity for $\omega>0$, and $A^{<}_{\mathrm{S\to H}}(k,\omega)$ has intensity for $\omega<0$.
    The calculations are performed by exact diagonalization method.
    Here, we use $U=20$, $\eta=0.15$, and $L=14$.
    } \label{fig:Spectra_ED}
\end{figure*}
\section{Numerical results of the photoemission spectrum}
\label{sec:Numerical_Results}

In this section, we show the photoemission spectrum $A(k,\omega)$ for photodoped Mott insulators evaluated with exact diagonalization based on the Lanczos method \cite{Avella_Strongly_2013}. 
Here, we approximate the delta function in the spectral function as $\delta(x) \approxeq -\frac{1}{\pi}\mathrm{Im}\left(\frac{1}{x+i\eta}\right)$ with $\eta$ being a small positive number (a broadening factor).
We set the system size to $L=14$.
In the following, we discuss three characteristic features that can be found in the photoemission spectrum of photodoped Mott insulators:
\begin{enumerate}
  \item A dispersive in-gap signal that appears even without the doublon-holon binding.
  \item A replica structure of the LHB signal that appears with the doublon-holon binding.
  \item Two flat band signals that appear below and above the LHB with the doublon-holon binding.
\end{enumerate}

Figure \hyperref[fig:Spectra_ED]{2(a)} shows the single-particle spectral function $A(k,\omega)$ for equilibrium Mott insulators ($N_d=N_h=0$).
Within the effective Hamiltonian \eqref{eq:Effective_Hamiltonian}, $V$ dependence does not appear in the $N_d=N_h=0$ sector.
The spectra exhibit characteristic features of Mott insulators, such as the LHB and UHB.
The former characterizes the occupied spectrum $A^{<}(k,\omega)$, while the latter characterizes the unoccupied spectrum $A^{>}(k,\omega)$.
The bandwidths of these bands are approximately $4t_{\mathrm{hop}}$, and the two bands are separated by the Mott gap of approximately $U-4t_{\mathrm{hop}}$.

Figures~\hyperref[fig:Spectra_ED]{2(b)}, \hyperref[fig:Spectra_ED]{2(c)}, and \hyperref[fig:Spectra_ED]{2(d)} show the occupied spectrum $A^{<}(k,\omega)$ for photodoped Mott insulators ($N_d=N_h=1$) with $V=0, 3, 5$, respectively.
For reference, we show the band edge of the UHB as the black dashed line.
Due to the photodoping, we can see that the in-gap spectra below the UHB emerge in $A^{<}_{\mathrm{D\to S}}(k,\omega)$.
Even when there is no doublon-holon binding ($V=0$), the in-gap spectrum shows a dispersive structure for $\abs{k} \lessapprox \pi/2$ (Fig.~\hyperref[fig:Spectra_ED]{2(b)}).
The dispersion width of the in-gap spectrum is in the same order as the spin exchange interaction $J$.
This is in stark contrast to the case of photodoped semiconductors, where the photoemission signal from unbound electrons and holes appears only at the bottom of the conduction band.
When we increase $V$, the in-gap spectrum shifts downwards, and the spectrum structure approaches that of the LHB.
This is analogous to the replica structure of the valence band observed in photodoped semiconductors with excitons.
Furthermore, with increasing $V$, we can also see that two flat bands above and below the LHB emerge in $A^{<}_{\mathrm{S\to H}}(k,\omega)$ (Fig.~\hyperref[fig:Spectra_ED]{2(d)}).
The energies of these flat bands are approximately $-U/2 \pm V$.
The intensity of the upper (lower) flat band around $k=\pm \frac{\pi}{2}$ ($k=\pm \pi$) is weaker than that in other momentum regions.
The momentum dependence of the flat band intensity is discussed in Appendix~\ref{sec:Trion_excitation_energy}.

\begin{figure*}[t!]
  \centering
  \includegraphics[width=0.9\textwidth]{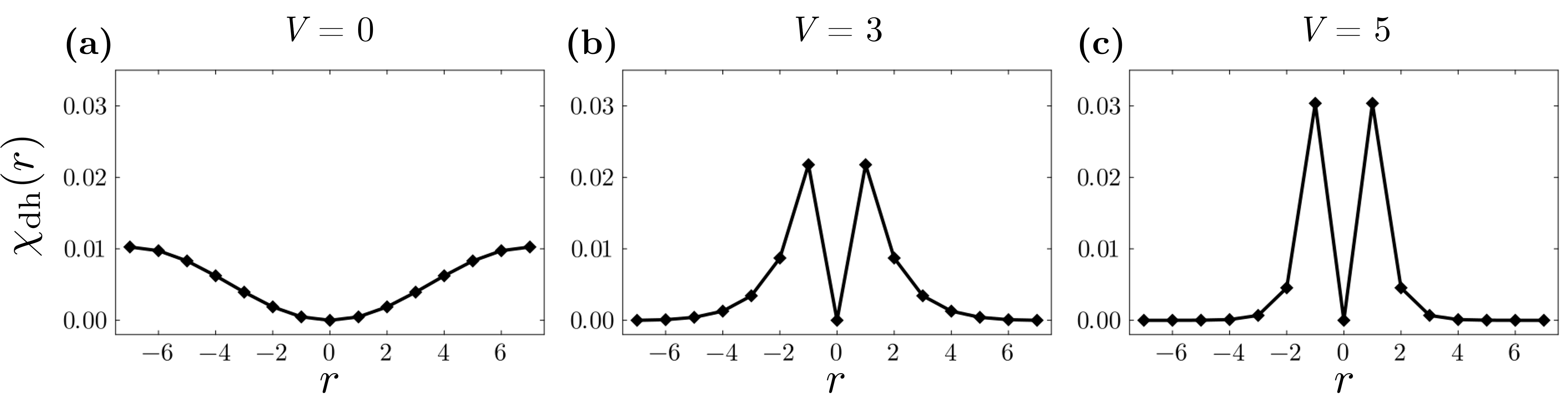}
  \caption{
    Doublon-holon correlation function $\chi_{\mathrm{dh}}(r)$ of photodoped Mott insulators ($N_d=N_h=1$) for (a) $V=0$, (b) $V=3$, and (c) $V=5$.
    The doublon and the holon are not bound for $V=0$, while they are bound for $V=3$ and $V=5$ (the binding becomes tighter as $V$ increases).
    The calculations are performed by the exact diagonalization method.
    Here we use $U=20$, and $L=14$. 
    } \label{fig:Correlations}
\end{figure*}

As shown in Fig.~\ref{fig:Spectra_ED}, the spectral features strongly depend on the nearest-neighbor interaction $V$.
This suggests that the photoemission spectrum is sensitive to the formation of the Mott-Hubbard exciton.
In order to confirm the exciton formation, we calculate the doublon-holon correlation function defined as
\begin{align}
  \chi_{\mathrm{dh}}(r) = \frac{1}{L}\sum_{j}\bra{\psi_0} \hat{n}^{d}_j \hat{n}^{h}_{j+r}\ket{\psi_0},
\end{align}
where $\hat{n}^{d}_j = \hat{n}_{j,\uparrow}\hat{n}_{j,\downarrow}$ is the doublon number operator, and $\hat{n}^{h}_j = (1-\hat{n}_{j,\uparrow})(1-\hat{n}_{j,\downarrow})$ is the holon number operator.
Figures \hyperref[fig:Correlations]{3(a)}, \hyperref[fig:Correlations]{3(b)}, and \hyperref[fig:Correlations]{3(c)} show $\chi_{\mathrm{dh}}(r)$ for $V=0, 3, 5$, respectively.
For $V=0$, $\chi_{\mathrm{dh}}(r)$ has the maximum at the largest distance $r=L/2$, suggesting that the doublon and the holon are not bound.
On the other hand, for $V=3$ and $5$, $\chi_{\mathrm{dh}}(r)$ has the maximum at the nearest-neighbor site ($r=1$), which means that the doublon and the holon are bound to form the Mott-Hubbard exciton.
The doublon-holon binding becomes tighter as $V$ increases.
Therefore, the appearance of the replica structure of the LHB and the two flat bands in Fig.~\ref{fig:Spectra_ED} should be closely related to the formation of the Mott-Hubbard exciton.

In order to clarify the relation between the spectral features and the formation of the Mott-Hubbard exciton, we investigate the spectral origin by two methods.
First, we study the dynamics of elementary excitations (holons, spinons) after the photoemission process using the time-dependent Lanczos method (Sec. \ref{sec:Time-dependent_Lanczos_method}).
In this approach, we can intuitively understand that the elementary excitation dynamics is qualitatively different depending on the presence or absence of exciton formation.
Second, we decompose the photoemission spectra into doublon, holon, and spinon spectra using the slave-particle method (Sec. \ref{sec:Slave-particle_method}).
This approach provides simplified yet direct insight into how each elementary excitation contributes to complex features in photoemission spectra.

Finally, we comment on the relation to previous studies.
First, the single-particle spectra of a doped one-dimensional Mott insulator in equilibrium have been studied in detail~\cite{Kohno_Spectral_2010}. 
The emergence of an in-gap dispersive band originating from spinons has been pointed out, and this feature appears similar to the photoemission signal from a photodoped Mott insulator without excitons. 
In general, however, in one-dimensional Mott systems, it is difficult to chemically dope carriers without destroying the chain structure. 
Thus, photodoping may provide an alternative route to studying such features.
Second, the photoemission spectrum of a photoexcited one-dimensional Mott insulator has previously been studied using pump-probe simulations~\cite{wang_Producing_2017,ejima_Nonequilibrium_2022,Sugimoto_Pumpprobe_2023}.
The photoemission spectra obtained within our steady-state approach are consistent with those after the pulse irradiation in these previous studies.
This indicates that our steady-state approach can capture the spectral features after the intraband relaxation process.
The main virtue of the steady-state formulation is its simplified treatment of photodoped states.
This simplification allows us to systematically analyze the photoemission spectral features and reveal their physical origin, as shown in the following sections.

\section{Elementary excitation dynamics} \label{sec:Time-dependent_Lanczos_method}

In this section, we investigate the dynamics of elementary excitations triggered by the photoemission process, to understand the origin of the characteristic spectral features. 
Practically, we study the time evolution from an initial state with one electron removed from the system, $\ket{\psi(t=0)} \propto \hat{c}_{i,\sigma}\ket{\psi_0}$.
This is because the occupied spectrum is nothing but the imaginary part of the lesser Green's function $G^<_{k,\sigma}(\omega)$, which is the spatial and temporal Fourier transform of
\begin{align}
  G^{<}_{ij,\sigma}(t) &= i \bra{\psi_0}\hat{c}^{\dagger}_{j,\sigma} \hat{c}_{i,\sigma}(t) \ket{\psi_0} \nonumber \\
  &= i\;e^{-iE_0t}\bra{\psi_0} \hat{c}^{\dagger}_{j,\sigma} e^{i\hat{H}_\mathrm{eff}t}\hat{c}_{i,\sigma}\ket{\psi_0}. \label{eq:Time_dependent_lesser_Green_function}
\end{align}
Namely, the occupied spectrum directly reflects the elementary excitation dynamics generated by the electron annihilation.

In practice, using the time-dependent Lanczos method \cite{Avella_Strongly_2013}, we simulate the (inverse) time-evolution process given by $\hat{H}_{\mathrm{eff}}$ as \begin{align}
  \ket{\psi(t)} = e^{i\hat{H}_{\mathrm{eff}} t} \ket{\psi(0)}. \label{eq:Time_evolution}
\end{align}
Remember that the photoemission process can be decomposed into two distinct processes as shown in Eq.~\eqref{eq:A_lesser2}: the doublon annihilation (singlon creation) process and the singlon annihilation (holon creation) process.
Reflecting this fact, we set $\ket{\psi(0)}$ as a projected state of $\hat{c}_{i,\sigma}\ket{\psi_0}$ to represent these photoemission processes.
In Sec.~\ref{sec:Doublon_annihilation_process}, we discuss the doublon annihilation process corresponding to $A^{<}_{\mathrm{D\to S}}(k,\omega)$. 
In Sec.~\ref{sec:Spin_annihilation_process}, we discuss the singlon annihilation process corresponding to $A^{<}_{\mathrm{S\to H}}(k,\omega)$.

To track the dynamics of the relevant elementary excitations, namely holons and spinons, we introduce the holon density $n^{\mathrm{holon}}(r,t)$ and the spinon density $n^{\mathrm{spinon}}(r,t)$ at position $r$ and time $t$ as
\begin{align}
  n^{\mathrm{holon}}(r,t) &= \bra{\psi(t)} \hat{n}^{h}_r \ket{\psi(t)}, \\
  n^{\mathrm{spinon}}(r,t) &= \bra{\psi(t)} \hat{n}^{s}_{r-1/2,\downarrow}\hat{n}^{s}_{r+1/2,\downarrow} \ket{\psi(t)}.
\end{align}
Here, $\hat{n}^{s}_{j,\sigma} = \hat{n}_{j,\sigma}(1-\hat{n}_{j,\bar{\sigma}})$ denotes the singlon number operator at site $j$ with spin $\sigma$.
We define the spinon position as the center of a ferromagnetic domain; accordingly, $r$ takes half-integer values.

\subsection{Doublon-annihilation (singlon-creation) process} \label{sec:Doublon_annihilation_process}

\begin{figure*}[t]
  \centering
  \includegraphics[width=0.9\textwidth]{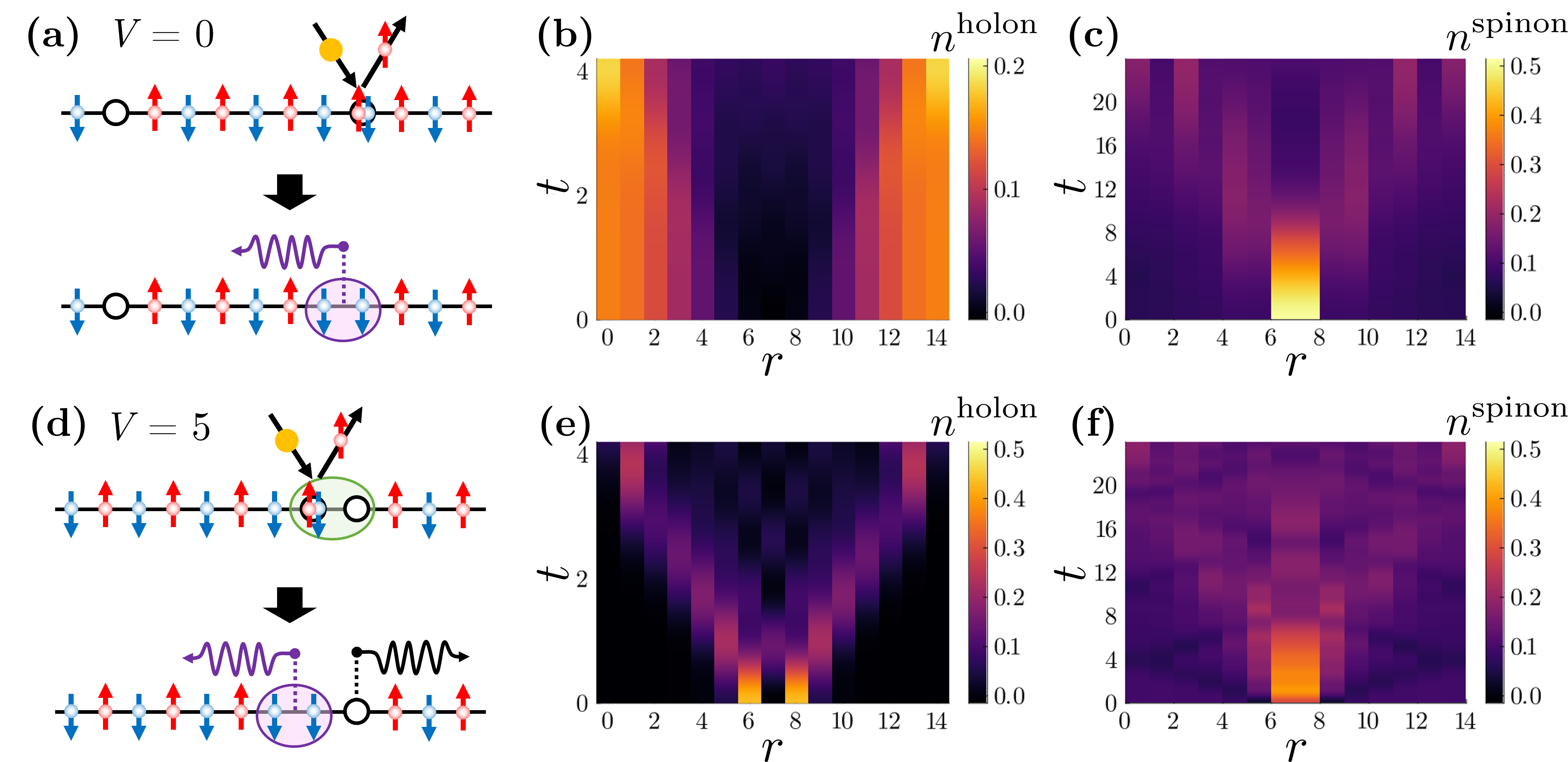}
  \caption{
    (a), (d) Schematic picture of the doublon annihilation process when (a) no exciton is formed and (d) an exciton is formed.
    When no exciton is formed, the photoemission process only creates the spinon (purple circle), and 
    the holon remains at the same position since the holon is far away from the spinon.
    When an exciton is formed, the photoemission process triggers both holon and spinon dynamics.
    (b), (e) Time evolution of the holon number $n^{\mathrm{holon}}(r,t)$ for (b) $V=0$ and (e) $V=5$, respectively.
    (c), (f) Time evolution of the spinon number $n^{\mathrm{spinon}}(r,t)$ for (c) $V=0$ and (f) $V=5$, respectively.
    The initial state is the one with a doublon annihilated at site $L/2$.
    The calculations are performed by the time-dependent Lanczos method.
    Here the parameters are $U=20$ and $L=14$.
    } \label{fig:Spectral_Origin}
\end{figure*}

To represent the doublon annihilation (or singlon creation) process, we take as the initial state the normalized state obtained by annihilating a doublon at site $L/2$;
\begin{align}
  \ket{\psi(t=0)}=
  \frac{\hat{c}_{L/2,\uparrow,\mathrm{D\to S}}\ket{\psi_0}}
  {\| \hat{c}_{L/2,\uparrow,\mathrm{D\to S}}\ket{\psi_0}\|}.
\end{align}
As shown below, the time evolution of this state reveals the following picture.
The doublon annihilation process inevitably generates spin dynamics (spinons), regardless of whether an exciton is formed, whereas the emergence of holon dynamics depends sensitively on exciton formation.
The behavior of the elementary excitations is consistent with the in-gap spectral features observed in $A^{<}_{\mathrm{D\to S}}(k,\omega)$.

First, we focus on the case without exciton formation.
To figure out the elementary excitation dynamics, it is helpful to consider a simple classical picture shown in Fig.~\hyperref[fig:Spectral_Origin]{4(a)}.
When no exciton is present, the doublon and holon are well separated.
In this case, photoemission annihilates the doublon while simultaneously creating a singlon.
This singlon generates a ferromagnetic domain wall in the antiferromagnetic background, corresponding to a spinon excitation.
The spinon can propagate due to the spin-exchange term in the Heisenberg Hamiltonian $\hat{H}_{\mathrm{s}}$ in Eq.~\eqref{eq:Effective_Hamiltonian}. 
In contrast, the holon remains localized because it is far away from the spinon.
The real-time evolution indeed confirms this picture.
Figures \hyperref[fig:Spectral_Origin]{4(b)} and \hyperref[fig:Spectral_Origin]{4(c)} show $n^{\mathrm{holon}}(r,t)$ and $n^{\mathrm{spinon}}(r,t)$, respectively.
We can see that photoemission creates a spinon at site $r=\frac{L}{2} \pm \frac{1}{2}$, while the holon is located at site $r=0=L$ when $t=0$.
As time evolves, the spinon propagates while the holon remains at $r=0=L$.
The motion of the spinon gives rise to the dispersive structure observed in Fig.~\hyperref[fig:Spectra_ED]{2(b)}.
The characteristic propagation time of the spinon is on the order of $1/J$, which is consistent with the spinon dispersion width $\sim J$.

Second, we focus on the case with an exciton, 
where the doublon and holon predominantly occupy neighboring sites [Fig.~\hyperref[fig:Spectral_Origin]{4(d)}].
Photoemission from a doublon creates a spinon, as in the case without an exciton, but at the same time releases the holon bound to the doublon, enabling its free propagation.
This picture is also confirmed by the real-time simulation.
Figures \hyperref[fig:Spectral_Origin]{4(e)} and \hyperref[fig:Spectral_Origin]{4(f)} show $n^{\mathrm{holon}}(r,t)$ and $n^{\mathrm{spinon}}(r,t)$, respectively.
We can see that photoemission creates a spinon at site $r=\frac{L}{2} \pm \frac{1}{2}$, while a holon is located at site $r=\frac{L}{2} \pm 1$ when $t=0$.
As time evolves, both the spinon and the holon propagate in the system.
Their coupled motion gives rise to the intricate in-gap structure in Figs.~\hyperref[fig:Spectra_ED]{2(c)} and \hyperref[fig:Spectra_ED]{2(d)}, which resembles the LHB. 
This similarity may seem natural, since the LHB also originates from the coupled motion of a spinon and a holon as we discuss in the next section.
The propagation time of the spinon is on the order of $1/J$, while the holon propagation time is on the order of $1/t_{\mathrm{hop}}$, which is consistent with the replica LHB dispersion width $\sim 4t_{\mathrm{hop}}$.

\subsection{Singlon-annihilation (holon-creation) process} \label{sec:Spin_annihilation_process}

\begin{figure*}[t]
  \centering
  \includegraphics[width=0.9\textwidth]{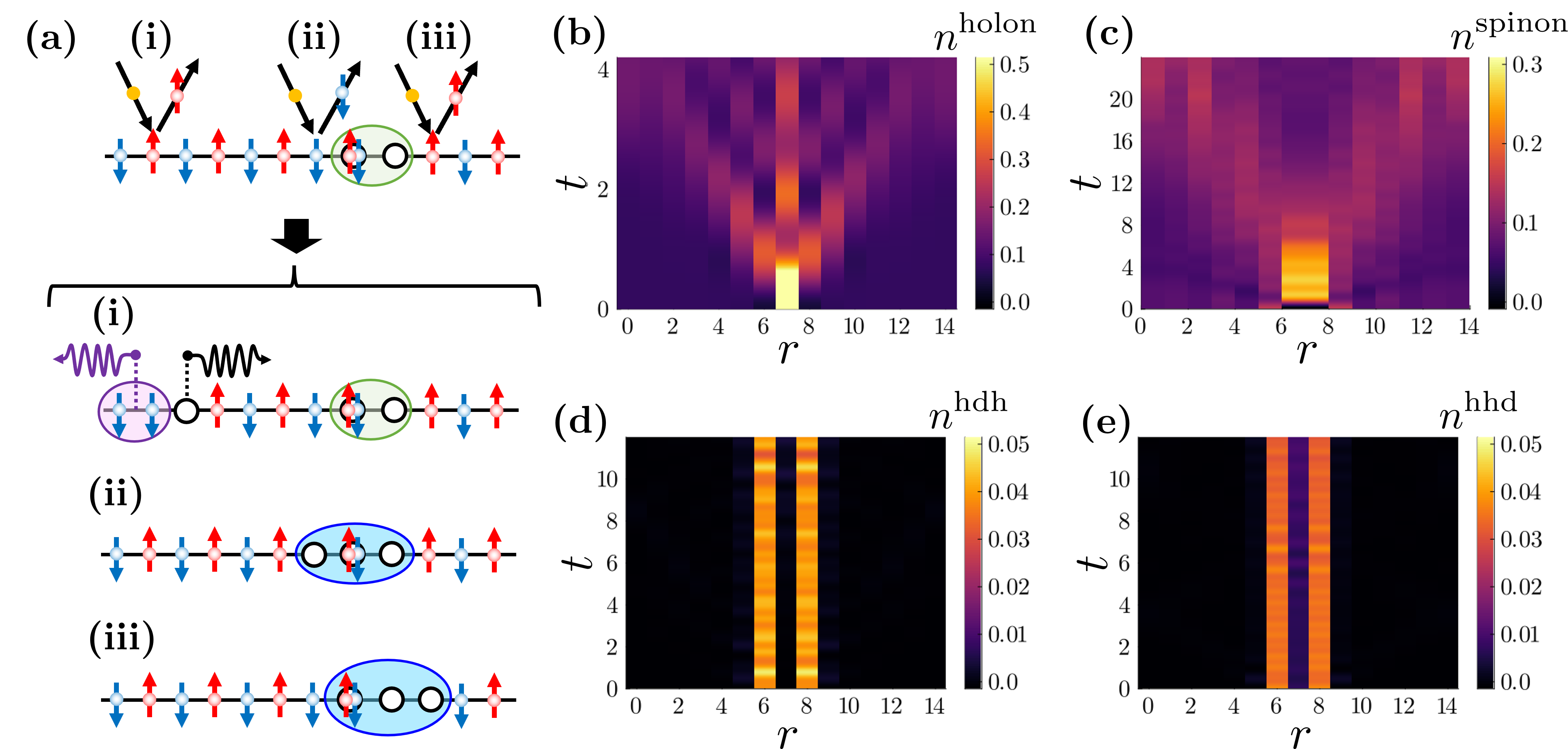}
  \caption{
    (a) Schematic picture of the singlon annihilation process in the presence of an exciton.
    There are three cases: (i) photoemission far from the exciton creating a holon and a spinon, (ii) photoemission near the exciton creating a holon-doublon-holon (hdh) trion, and (iii) photoemission near the exciton creating a holon-holon-doublon (hhd) trion.
    (b), (c), (d), (e) Time evolution of (b) the holon number $n^{\mathrm{holon}}(r,t)$, (c) the spinon number $n^{\mathrm{spinon}}(r,t)$, (d) the hdh trion number $n^{\mathrm{hdh}}(r,t)$, and (e) the hhd trion number $n^{\mathrm{hhd}}(r,t)$.
    The initial state is obtained after singlon annihilation at site $L/2$.
    The calculations are performed by the time-dependent Lanczos method.
    The parameters are $U=20$, $V=5$, and $L=14$.
    } \label{fig:Spectral_Origin_flat}
\end{figure*}

To represent the singlon-annihilation (holon-creation) process, we set the initial state to the one obtained after singlon annihilation at site $L/2$;
\begin{align} \ket{\psi(t=0)}=\frac{\hat{c}_{L/2,\uparrow,\mathrm{S\to H}}\ket{\psi_0}}{\| \hat{c}_{L/2,\uparrow,\mathrm{S\to H}}\ket{\psi_0}\|}.
\end{align}
The time evolution of this state confirms the following picture (Fig.~\hyperref[fig:Spectral_Origin_flat]{5(a)}).
After this photoemission process, two holons and one doublon remain in the system.
When the singlon is annihilated far from the preexisting holon and doublon (or exciton, if present), both holon and spinon excitations are generated, giving rise to the LHB.
The LHB of the equilibrium (undoped) Mott insulator originates from a similar process.
In the presence of an exciton, however, singlon annihilation can also occur in its vicinity, leading to the formation of two types of trions, i.e., bound states of two holons and one doublon.
One is the holon-doublon-holon (hdh) trion, and the other is the holon-holon-doublon (hhd) trion.
Within the semiclassical picture, the energy difference between the hdh (hhd) trion state and the state before photoemission is estimated as $\sim -U/2 + V$ ($\sim -U/2 - V$).
Therefore, the hdh (hhd) trion spectrum appears above (below) the LHB.
Once formed, the trion hardly propagates, because the Hamiltonian does not contain a direct trion-hopping term.

These features are indeed confirmed by following the time evolution of the holon number, the spinon number, the hdh trion number $n^{\mathrm{hdh}}(r,t)$, and the hhd trion number $n^{\mathrm{hhd}}(r,t)$.
The hdh trion number and the hhd trion number are defined as
\begin{align}
  n^{\mathrm{hdh}}(r,t) &= \bra{\psi(t)} \hat{n}^{h}_{r-1}\hat{n}^{d}_{r}\hat{n}^{h}_{r+1} \ket{\psi(t)}, \\
  n^{\mathrm{hhd}}(r,t) &= \bra{\psi(t)} (\hat{n}^{h}_{r-1}\hat{n}^{h}_{r}\hat{n}^{d}_{r+1} + \hat{n}^{d}_{r-1}\hat{n}^{h}_{r}\hat{n}^{h}_{r+1}) \ket{\psi(t)}.
\end{align}
Here we define the trion position as the center site of the trion.
Figures \hyperref[fig:Spectral_Origin_flat]{5(b)} and \hyperref[fig:Spectral_Origin_flat]{5(c)} show $n^{\mathrm{holon}}(r,t)$ and $n^{\mathrm{spinon}}(r,t)$, respectively.
We can see that photoemission creates a holon at site $r=L/2$ when $t=0$, and a spinon is created at site $r=\frac{L}{2} \pm \frac{1}{2}$ after the holon propagation.
As time evolves, both the holon and the spinon propagate in the system, resulting in the LHB spectrum observed in Figs.~\hyperref[fig:Spectra_ED]{2(b)}-\hyperref[fig:Spectra_ED]{2(d)}.
Figures \hyperref[fig:Spectral_Origin_flat]{5(d)}, and \hyperref[fig:Spectral_Origin_flat]{5(e)} show $n^{\mathrm{hdh}}(r,t)$ and $n^{\mathrm{hhd}}(r,t)$, respectively.
We can see that both hdh and hhd trions are created at $r=\frac{L}{2} \pm 1$ when $t=0$.
As time evolves, both the hdh and hhd trions stay at the same position.
The nearly-immobile behavior of the trions results in the flat band spectra observed in Fig.~\hyperref[fig:Spectra_ED]{2(d)}.

\section{Slave-particle approach} \label{sec:Slave-particle_method}

In this section, we investigate the relation between the photoemission spectra and spectral functions of elementary excitations using the slave-particle approach.
Within the slave-particle representation, the electron operators are decomposed into holon, doublon, and spinon operators. 
Using this formulation, we can approximately decompose the electron spectral function into those associated with each elementary excitation \cite{Sorella_Onehole_1992,Penc_Spectral_1995,Penc_Spectral_1997,Penc_Finitetemperature_1997,Suzuura_Spincharge_1997,Bohrdt_Angleresolved_2018}.
This analysis shows that the in-gap spectrum originates directly from the gapless spinon dispersion when no exciton is formed. 
It also demonstrates that the replica LHB reflects the spatial extent of the Mott-Hubbard exciton wavefunction. 

\subsection{Formulation}

First, we review the slave-particle representation~\cite{Wen_MeanField_2007, Baskaran_Resonating_1987, Huang_spin_2023}.
In this formalism, we express the electron operator as 
\begin{align}
    \hat{c}_{j,\sigma} = \hat{h}^{\dagger}_j \hat{f}_{j,\sigma} + (-1)^{\sigma} \hat{d}_j\hat{f}^{\dagger}_{j,\bar{\sigma}}, \label{eq:Slave_particle_representation}
\end{align}
where $\hat{h}_j$ ($\hat{d}_j$) is the holon (doublon) annihilation operator at site $j$, and $\hat{f}_{j,\sigma}$ is 
the spinon annihilation operator with spin $\sigma$ at site $j$.
Note that $\hat{h}_j$ and $\hat{d}_j$ are bosonic operators, while $\hat{f}_{j,\sigma}$ is a fermionic operator.
These operators satisfy the commutation relations, $[\hat{h}_i,\hat{h}^{\dagger}_j] = \delta_{i,j}$, $[\hat{d}_i,\hat{d}^{\dagger}_j] = \delta_{i,j}$, and the anticommutation relation, $\{\hat{f}_{i,\sigma},\hat{f}^{\dagger}_{j,\sigma'}\} = \delta_{i,j}\delta_{\sigma,\sigma'}$.
The other commutation relations are zero.
Equation \eqref{eq:Slave_particle_representation} is consistent with the anticommutation relation of the electron operators $\{\hat{c}_{i,\sigma},\hat{c}^{\dagger}_{j,\sigma'}\} = \delta_{i,j}\delta_{\sigma,\sigma'}$.

\begin{figure}[t!]
  \centering
  \includegraphics[width=0.483\textwidth]{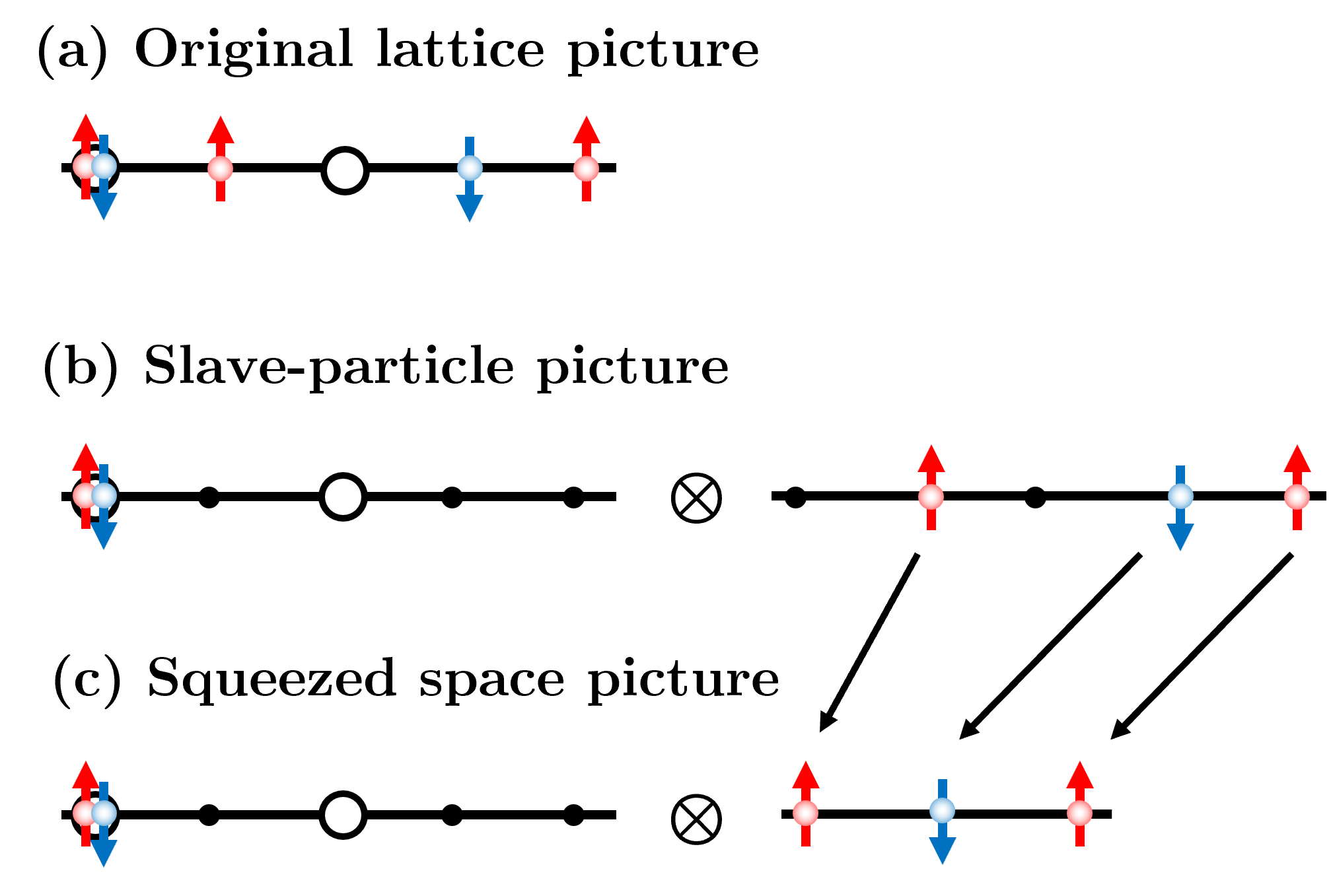}
  \caption{
    (a) Schematic picture of a state $\ket{d,\uparrow, h, \downarrow, \uparrow}$ in the original electron basis.
    (b) The corresponding slave-particle representation $ \hat{d}^{\dagger}_1 \hat{h}^{\dagger}_3 \ket{0_c}\otimes \hat{f}^{\dagger}_{2,\uparrow} \hat{f}^{\dagger}_{4,\downarrow}\hat{f}^{\dagger}_{5,\uparrow}\ket{0_s}$.
    (c) The corresponding squeezed-space representation $ \hat{d}^{\dagger}_1 \hat{h}^{\dagger}_3 \ket{0_c}\otimes \hat{Z}^{\dagger}_{1,\uparrow} \hat{Z}^{\dagger}_{2,\downarrow}\hat{Z}^{\dagger}_{3,\uparrow}\ket{0_s}$.
    } \label{fig:SlaveParticle}
\end{figure}

The first term in Eq.~\eqref{eq:Slave_particle_representation} creates a holon and annihilates a spinon, which corresponds to $\hat{c}_{j,\sigma,\mathrm{S\to H}}$ in Eq.~\eqref{eq:c_StoH}.
The second term in Eq.~\eqref{eq:Slave_particle_representation} annihilates a doublon and creates a spinon, which corresponds to $\hat{c}_{j,\sigma,\mathrm{D\to S}}$ in Eq.~\eqref{eq:c_DtoS}.
Since every site is either empty, singly occupied, or doubly occupied, we need to impose the following local constraint for every site $j$:
\begin{align}
    \hat{d}^{\dagger}_j \hat{d}_j + \sum_{\sigma} \hat{f}^{\dagger}_{j,\sigma} \hat{f}_{j,\sigma} + \hat{h}^{\dagger}_j \hat{h}_j = 1. 
    \label{eq:Local_constraint}
\end{align}
The slave-particle representation allows us to express the charge and spin degrees of freedom separately.
For example, the original electron basis $\ket{\mathrm{d},\uparrow,\mathrm{h},\downarrow,\uparrow}$, which corresponds to a doublon at site 1, a holon at site 3, spin-up singlons at site 2 and 5, and a spin-down singlon at site 4 (Fig. \hyperref[fig:SlaveParticle]{4(a)}), can be expressed in the slave-particle representation as $ \hat{d}^{\dagger}_1 \hat{h}^{\dagger}_3 \ket{0_c}\otimes \hat{f}^{\dagger}_{2,\uparrow} \hat{f}^{\dagger}_{4,\downarrow}\hat{f}^{\dagger}_{5,\uparrow}\ket{0_s}$ (Fig. \hyperref[fig:SlaveParticle]{4(b)}).
Here, $\ket{0_c}$ ($\ket{0_s}$) denotes the vacuum state of the charge (spin) part.

To further simplify the slave particle representation, we introduce a new spin basis called the squeezed spin space~\cite{Ogata_Betheansatz_1990,Murakami_Spin_2023}.
In the squeezed spin space, all empty and doubly occupied sites are removed from the spin basis, so that only singly-occupied sites remain, and the site indices of these singly-occupied sites are relabeled.
The relabeling is implemented through the creation operators of the squeezed space as $\hat{Z}^{\dagger}_{l(j),\sigma} = \hat{f}^{\dagger}_{j,\sigma}$, where $l(j) = j-\sum_{m<j}(\hat{d}^{\dagger}_m\hat{d}_m + \hat{h}^{\dagger}_m\hat{h}_m)$ denotes the site index in the squeezed space.
In the squeezed-space representation, the local constraint \eqref{eq:Local_constraint} is automatically satisfied as long as the doublon and holon operators obey the hard-core conditions.
The information of the local constraint is encoded in the squeezed space lattice structure $l(j)$.
For example, the previous example $\ket{\mathrm{d},\uparrow,\mathrm{h},\downarrow,\uparrow}$ can be expressed in the squeezed space representation as $ \hat{d}^{\dagger}_1 \hat{h}^{\dagger}_3 \ket{0_c}\otimes \hat{Z}^{\dagger}_{1,\uparrow} \hat{Z}^{\dagger}_{2,\downarrow}\hat{Z}^{\dagger}_{3,\uparrow}\ket{0_s}$ (Fig.~\hyperref[fig:SlaveParticle]{4(c)}).

In the squeezed space slave-particle representation, the effective Hamiltonian \eqref{eq:Effective_Hamiltonian} is rewritten as
\begin{align}
  \hat{H}_{\mathrm{kin,h}} &= -t_{\mathrm{hop}} \sum_{j} \hat{h}^{\dagger}_{j} \hat{h}_{j+1} + \mathrm{h.c.}, \label{eq:H_kin_d} \\
  \hat{H}_{\mathrm{kin,d}} &= t_{\mathrm{hop}} \sum_{j} \hat{d}^{\dagger}_{j} \hat{d}_{j+1} + \mathrm{h.c.}, \label{eq:H_kin_h} \\
  \hat{H}_{\mathrm{U}} &= \frac{U}{2} \sum_{j} \qty(\hat{n}^{\mathrm{d}}_j + \hat{n}^{\mathrm{h}}_j), \label{eq:H_U} \\
  \hat{H}_{\mathrm{V}} + \hat{H}_{\mathrm{\eta}} &= (V-\frac{J}{4}) \sum_{j} \qty(\hat{n}^{\mathrm{d}}_j - \hat{n}^{\mathrm{h}}_j)(\hat{n}^{\mathrm{d}}_{j+1} - \hat{n}^{\mathrm{h}}_{j+1}) \nonumber \\
  &\quad + \frac{J}{2} \sum_{j} \hat{d}^{\dagger}_j \hat{h}_j \hat{h}^{\dagger}_{j+1} \hat{d}_{j+1} + \mathrm{h.c.}, \label{eq:H_V_dh_ex} \\
  \hat{H}_{\mathrm{s}} &= J \sum_{l \in \mathrm{sq}} \delta_{j(l)+1,j(l+1)} \hat{\vb{s}}_l \cdot \hat{\vb{s}}_{l+1}, 
  \label{eq:H_spin_ex}
\end{align}
where $\hat{s} = \frac{1}{2}\sum_{\alpha, \beta} \hat{Z}^{\dagger}_{\alpha}\sigma_{\alpha,\beta}\hat{Z}_{\beta}$ is the spin operator in the squeezed space, $\hat{n}^{\mathrm{d}}_j = \hat{d}^{\dagger}_j \hat{d}_j$ is the doublon number operator, and $\hat{n}^{\mathrm{h}}_j = \hat{h}^{\dagger}_j \hat{h}_j$ is the holon number operator.
In the spin-spin interaction Hamiltonian $\hat{H}_{\mathrm{s}}$, the summation is taken over the squeezed space site index $l$.
This means that the spin-spin interaction exists only when the two spins occupy nearest-neighboring sites in the original lattice (i.e., $j(l)+1=j(l+1)$).
The main advantage of using the squeezed space representation is that the hopping term does not depend on the spin configuration (see Appendix~\ref{sec:Slave_particle_squeezed_space} for details).
Instead, the spin-spin interaction term depends on the doublon and holon configuration.

While the reformulation presented so far is exact, the coupling between the charge and spin degrees of freedom makes the problem difficult to analyze. 
Therefore, in the following, we introduce an approximation that decouples these degrees of freedom.
We consider the situation where the density of doublons and holons is low, which is precisely the regime of our interest.
In this situation, the probability that two spins occupy nearest-neighbor sites is close to unity.
Then, the spin-spin interaction term can be approximated by the usual Heisenberg Hamiltonian in the squeezed lattice:
\begin{align}
  \hat{H}_{\mathrm{s}} \approx J \sum_{l \in \mathrm{sq}} \vb{s}_l \cdot \vb{s}_{l+1}. \label{eq:H_spin_ex_approx}
\end{align}
Then, the total Hamiltonian can be approximately separated into the charge part $\hat{H}_{\mathrm{c}}$ and the spin part $\hat{H}_{\mathrm{s}}$.
The ground-state energy and wavefunction are also approximately factorized into the charge and spin parts \cite{Ogata_Betheansatz_1990,Murakami_Spin_2023}.
We denote the eigenenergies and eigenstates of the charge part by $E^{\mathrm{c}}_n$ and $\ket{c_n}$, and those of the spin part by $E^{\mathrm{s}}_n$ and $\ket{s_n}$.
Then, the eigenenergies and eigenstates of the total system are approximately given by $E_n = E^{\mathrm{c}}_m + E^{\mathrm{s}}_l$ and $\ket{\psi_n} = \ket{c_m} \otimes \ket{s_l}$.
At low doublon and holon densities, we may also approximate the slave-particle representation of the electron operators in momentum space as
\begin{align}
    \hat{c}_{k,\sigma} \approx& \frac{1}{\sqrt{L}}\sum_{k_s} \qty(\hat{h}^{\dagger}_{k_s - k} \hat{Z}_{k_s,\sigma} + (-1)^{\sigma} \hat{d}_{k_s + k} \hat{Z}^{\dagger}_{k_s,\bar{\sigma}}). \label{eq:Slave_particle_representation_momentum}
\end{align}
Here $\hat{d}^{\dagger}_k$, $\hat{h}^{\dagger}_k$, and $\hat{Z}^{\dagger}_{k,\sigma}$ are the Fourier transforms of $\hat{d}^{\dagger}_j$, $\hat{h}^{\dagger}_j$, and $\hat{Z}^{\dagger}_{j,\sigma}$, respectively.
An argument leading to the approximate relation \eqref{eq:Slave_particle_representation_momentum} is given in Appendix~\ref{sec:Spectra_decomposition}.
In this representation, the momentum of the annihilated electron is shared by the charge and spin parts.

Under the decoupling approximation, we find the simple decomposition formulas for the occupied spectral functions Eqs.~\eqref{eq:A_lesser_DtoS} and \eqref{eq:A_lesser_StoH} as
\begin{align}
  A^{<}_{\mathrm{D\to S}}(k,\omega) &= \frac{1}{L} \sum_{k_s,k_c}\int d\omega_c d\omega_s \nonumber \\
  &\quad\times \delta_{k,k_c-k_s} \delta(\omega - (\omega_c - \omega_s)) \nonumber \\
  &\quad\times A^{<}_d(k_c,\omega_c) A^{>}_{Z}(k_s,\omega_s), \label{eq:Spectral_decomposition1} \\
  A^{<}_{\mathrm{S\to H}}(k,\omega) &= \frac{1}{L}\sum_{k_s,k_c}\int d\omega_c d\omega_s \nonumber \\
  &\quad\times \delta_{k,k_s-k_c} \delta(\omega - (\omega_s - \omega_c)) \nonumber \\
  &\quad\times A^{>}_h(k_c,\omega_c) A^{<}_{Z}(k_s,\omega_s), \label{eq:Spectral_decomposition2} 
\end{align}
where 
\begin{align}
  A^{<}_d(k_c,\omega_c) =& \sum_m \delta(\omega_c + E^{\mathrm{c}}_{m0})\abs{\bra{c_m}\hat{d}_{k_c}\ket{c_0}}^2 , \\
  A^{>}_h(k_c,\omega_c) =& \sum_m \delta(\omega_c - E^{\mathrm{c}}_{m0})\abs{\bra{c_m}\hat{h}^{\dagger}_{k_c}\ket{c_0}}^2 , \\
  A^{<}_{Z}(k_s,\omega_s) =& \sum_m \delta(\omega_s + E^{\mathrm{s}}_{m0})\abs{\bra{s_m}\hat{Z}_{k_s,\sigma}\ket{s_0}}^2 , \\
  A^{>}_{Z}(k_s,\omega_s) =& \sum_m \delta(\omega_s - E^{\mathrm{s}}_{m0})\abs{\bra{s_m}\hat{Z}^{\dagger}_{k_s,\sigma}\ket{s_0}}^2
\end{align}
are the momentum-resolved spectra of each slave particle. 
The derivation is given in Appendix~\ref{sec:Spectra_decomposition}.
Each slave particle spectrum reflects the elementary excitations of the charge and spin degrees of freedom.
These expressions in Eqs.~\eqref{eq:Spectral_decomposition1} and \eqref{eq:Spectral_decomposition2} tell us that the occupied spectrum is given by the convolution of the doublon (holon) spectrum and the spinon spectrum.

Before analyzing the slave particle spectra, we explain the physical meaning of the convolution integral in Eqs.~\eqref{eq:Spectral_decomposition1} and \eqref{eq:Spectral_decomposition2}.
Each slave particle should satisfy the conservation laws of momentum and energy.
Electron annihilation at a doubly occupied site with momentum $k$ and energy $\omega$ is composed of the doublon slave-particle annihilation with momentum $k_c$ and energy $\omega_c$ and the spinon slave-particle creation with momentum $k_s$ and energy $\omega_s$.
Due to the conservation laws, $k = k_c - k_s$ and $\omega = \omega_c - \omega_s$ hold (note that the signs of $k_s$ and $\omega_s$ are negative because a spinon is created).
Electron annihilation at a singly occupied site with momentum $k$ and energy $\omega$ is composed of the holon slave-particle creation with momentum $k_c$ and energy $\omega_c$ and the spinon slave-particle annihilation with momentum $k_s$ and energy $\omega_s$.
Due to the conservation laws, $k = k_s - k_c$ and $\omega = \omega_s - \omega_c$ hold (note that the signs of $k_c$ and $\omega_c$ are negative because a holon is created).

In the following subsection, we consider the slave-particle spectra one by one.
We can split the problem into the charge and spin parts.
In Sec.~\ref{sec:Charge_part}, we analyze the charge part problem with one doublon and one holon on the $L$-site lattice with an attractive interaction.
In Sec.~\ref{sec:Spin_part}, we analyze the spin part problem with $L-2$ spins on the $L-2$-site lattice with the Heisenberg interaction.
In Sec.~\ref{sec:Convolution_part}, we obtain the photoemission spectrum by performing the convolution integral [Eqs.~\eqref{eq:Spectral_decomposition1} and \eqref{eq:Spectral_decomposition2}] of slave particle spectra. 
We set the parameters to $U=20$ and $L=100$ for the following calculations.

\subsection{Spectrum of the doublon and holon slave particles} \label{sec:Charge_part}

\begin{figure}[t!]
  \centering
  \includegraphics[width=0.45\textwidth]{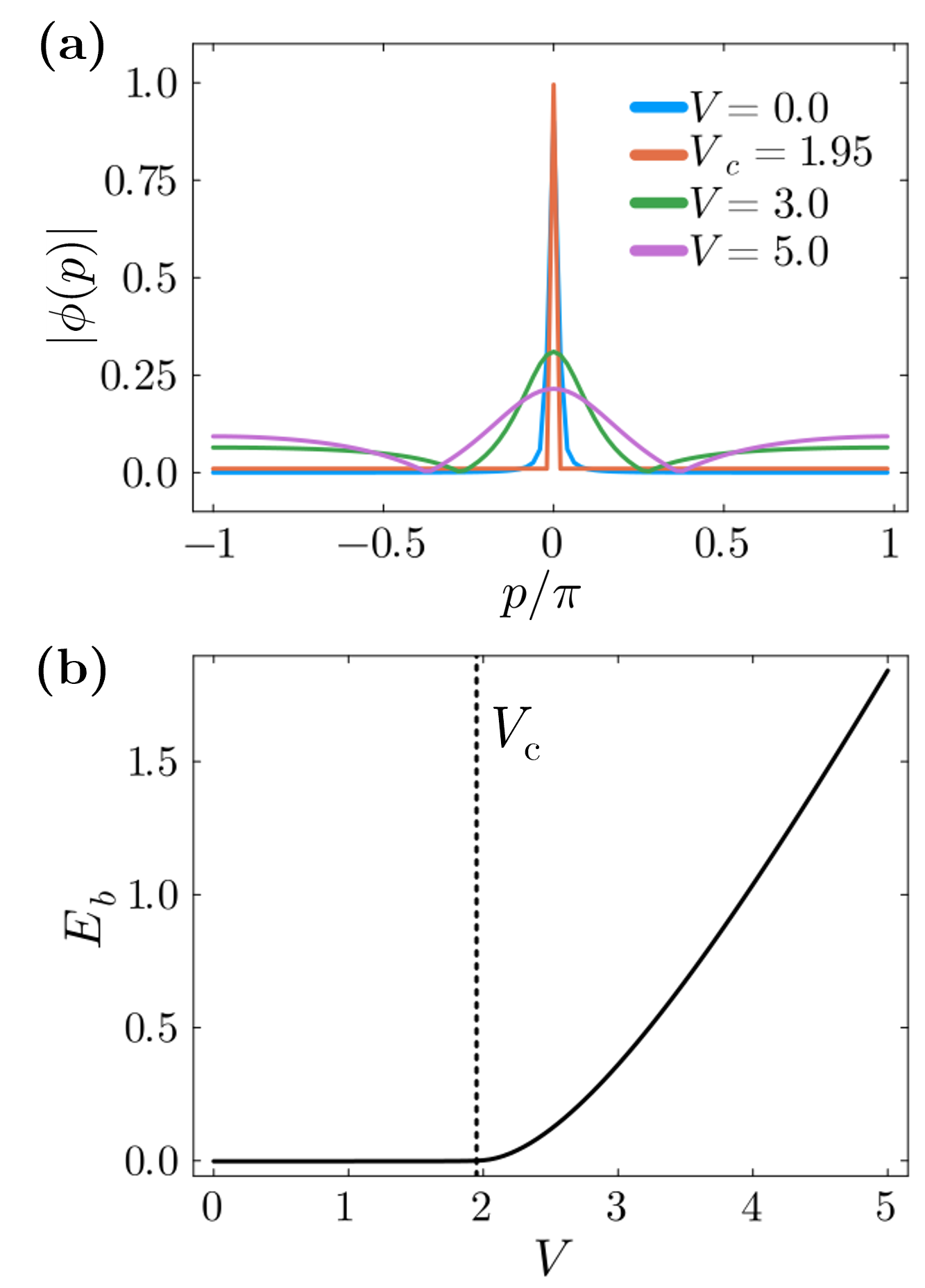}
  \caption{
    (a) Absolute value of the doublon-holon wavefunction in momentum space, $\abs{\phi(p)}$, for $V=0.0,~ 1.95,~ 3.0,~ 5.0$.
    Here $V_{\rm c}=1.95$ corresponds to the critical value for the formation of the doublon-holon binding.
    (b) The binding energy $E_b (= U-4t_{\rm hop}-E^c_0)$ as a function of $V$. The vertical dashed line indicates $V=V_{\rm c}$.
    The parameters are $U=20$ and $L=100$.
    } \label{fig:Charge_part_wave_Eb}
\end{figure}

\begin{figure*}[t!]
  \centering
  \includegraphics[width=0.9\textwidth]{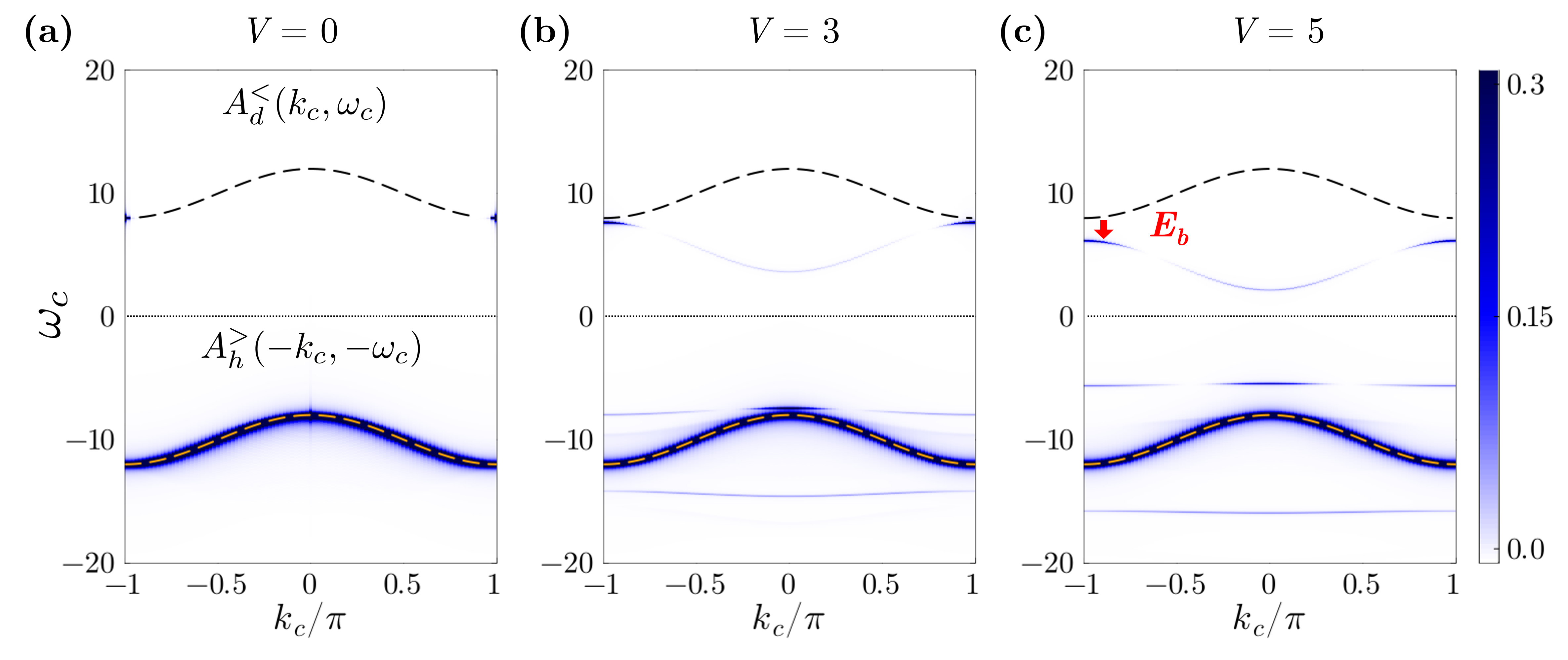}
  \caption{
    Occupied spectra of the doublon slave particle $A^{<}_d(k_c,\omega_c)$ and the unoccupied spectra of the holon slave particle $A^{>}_h(-k_c,-\omega_c)$ for (a) $V=0$, (b) $V=3$, and (c) $V=5$.
    The black dashed lines show the doublon band $\epsilon_d(k_c)$, and the orange dashed lines show the holon band $-\epsilon_h(-k_c)$. 
    The calculation is performed based on the exact diagonalization method.
    The parameters are $U=20$, and $L=100$.
    } \label{fig:Charge_part_spectra}
\end{figure*}

In this subsection, we analyze the spectra of the doublon and holon slave particles, i.e. $A^<_d(k_c,\omega_c)$ and $A^>_h(k_c,\omega_c)$.
We first derive the exact wavefunction of the state with one doublon and one holon described by the Hamiltonian $\hat{H}_c$ using the Bethe ansatz.
The wavefunction and the binding energy qualitatively change at the critical value of the attractive interaction for the formation of the doublon-holon binding.
We find that the energy of the doublon slave particle spectrum $A^<_d(k_c,\omega_c)$ reflects the doublon-holon binding energy, and the intensity distribution reflects the wavefunction.
On the other hand, the holon slave particle spectrum $A^>_h(k_c,\omega_c)$ reflects the trion formation, in which the flat band spectrum appears above and below the holon band.

First, we consider the dispersion relation of a free charge carrier (a doublon or a holon).
If a single doublon (or a single holon) exists in the system, the dispersion relations are given by 
\begin{align}
  \epsilon_d(k) =& 2t_{\mathrm{hop}}\cos(k) + U/2, \label{eq:doublon_band} \\
  \epsilon_h(k) =& -2t_{\mathrm{hop}}\cos(k) + U/2. \label{eq:holon_band}
\end{align}
The constant $U/2$ denotes the on-site energy per one doublon or one holon.
It is interesting to compare these dispersion relations with those of electrons and holes in indirect-gap semiconductors,
which typically have dispersion relations of
\begin{align}
  \epsilon_c(k) &= 2t_{\mathrm{hop}}\cos(k) + \Delta/2, \label{eq:conduction_band} \\
  \epsilon_v(k) &= 2t_{\mathrm{hop}}\cos(k) -\Delta/2.
  \label{eq:valence_band}
\end{align}
Since the hole is defined by annihilation of an electron in the valence band, the hole energy is given by $-\epsilon_v(-k)$.
Hence we have the correspondence between the doublon-holon and electron-hole bands as $\epsilon_d(k) \leftrightarrow \epsilon_c(k)$, $-\epsilon_h(-k) \leftrightarrow \epsilon_v(k)$.
In the following, we call $\epsilon_d(k)$ the doublon band and $-\epsilon_h(-k)$ the holon band.

Next, we consider the one doublon-holon ground state, which is generally written as 
\begin{align}
\ket{c_0} = \sum_{l,m}\Psi(l,m) \hat{d}^{\dagger}_{l}\hat{h}^{\dagger}_{m}\ket{0_c}, 
\end{align}
where $l$ ($m$) is the position of a doublon (holon).
Due to the hard-core boson nature of the doublon and holon, $\Psi(l,l)=0$ holds.
We derive an analytical form of the ground state wavefunction $\Psi(l,m)$ and the corresponding ground state energy $E^c_0$ for $\hat{H}_c$ (see Appendix~\ref{sec:Exact_dh_solution} for the derivation details).
Note that the exact wavefunction for $J=0$ has been obtained in Ref.~\cite{gallagher_Excitons_1997}.
We use the standard method to solve two-particle problems on a lattice~\cite{Giamarchi_Quantum_2003}.
The behavior of the ground state and energy qualitatively changes depending on the effective doublon-holon attractive interaction, $V_{\mathrm{eff}} = V + J/4$.
We also introduce the critical value of $V$ for the exciton formation as $V_{\rm c} = 2t_{\rm hop}-J/4$.

When $V_{\mathrm{eff}} < 2t_{\mathrm{hop}}$ ($V<V_{\rm c}$), the ground state is the scattering state of the doublon and holon,
\begin{align}
  \Psi(l,m) &\propto \cos(\pi l) \sin(k|m-l|+\varphi), \label{eq:dh_scattering_wf} \\
  E^c_0 &= U - 4t_{\mathrm{hop}}\cos(k), \label{eq:dh_scattering_energy}
\end{align}
where $K=\pi$ is the center of mass momentum, $k$ is the relative momentum of the doublon and holon, and $\varphi$ is the phase shift due to the doublon-holon interaction.
$k$ and $\varphi$ satisfy the following Bethe-ansatz equations:
\begin{align}
  e^{2i\varphi} &= \frac{V_{\mathrm{eff}}e^{-ik}-2t_{\mathrm{hop}}}{V_{\mathrm{eff}}e^{ik}-2t_{\mathrm{hop}}}, \\
  k &= \frac{2\pi\lambda+\pi-2\varphi}{L} \quad (\lambda = -\frac{L}{2},\dots,\frac{L}{2}-1).
\end{align}
If we neglect the $1/L$ correction, the ground state relative momentum is given by $k=0$, and the ground state energy is given by $E^c_0 = U - 4t_{\mathrm{hop}}$.

When $V_{\mathrm{eff}} > 2t_{\mathrm{hop}}$ ($V>V_{\rm c}$), the ground state is the bound state of the doublon and holon, 
\begin{align}
  \Psi(l,m) &\propto \cos(\pi l) e^{-\kappa |m-l|} , \label{eq:dh_bound_wf} \\
  E^c_0 &= U - V_{\mathrm{eff}} - \frac{4t_{\mathrm{hop}}^2}{V_{\mathrm{eff}}}, \label{eq:dh_bound_energy}
\end{align}
with $\kappa = \ln(V_{\mathrm{eff}}/2t_\text{hop})$.
The probability of the doublon and holon being close to each other increases as $V_{\mathrm{eff}}$ increases.
Since $V_{\mathrm{eff}} + 4t_{\mathrm{hop}}^2/V_{\mathrm{eff}} > 4t_{\mathrm{hop}}$ holds, the bound state energy appears below the edge of the scattering state continuum, $U-4t_{\mathrm{hop}}$.

In order to see the momentum-space structure of the doublon-holon wavefunction, we perform the Fourier transformation,
\begin{align}
  \phi(p) = \sum_{l,r} e^{i\pi l}e^{ipr} \Psi(l,l+r).
\end{align}
Using this wavefunction, we can rewrite the ground state as 
\begin{align}
\ket{c_0} = \sum_{p} \phi(p) \hat{d}^{\dagger}_{\pi-p}\hat{h}^{\dagger}_{p}\ket{0_c}.
\end{align}
We plot $\abs{\phi(p)}$ and the binding energy $E_b = U-4t_{\mathrm{hop}}-E^c_0$ in Figs.~\hyperref[fig:Charge_part_wave_Eb]{7(a)} and \hyperref[fig:Charge_part_wave_Eb]{7(b)}, respectively.
For $V < V_c$, $\abs{\phi(p)}$ has a sharp peak at $p=0$, meaning that the doublon and holon have the momenta $\pi$ and $0$, respectively.
In this case, the doublon and holon are almost independent particles.
The binding energy vanishes in this region. 
For $V > V_c$, $\abs{\phi(p)}$ has a broad distribution around $p=0$.
In this case, the doublon and holon are bound to each other.
In other words, the holon is localized around the doublon in real space.
There is also a broad distribution around $p=\pi$ due to the hard-core condition. 
The binding energy increases as $V$ increases.

Using $\phi(p)$ and the binding energy, we can write down $A^{<}_d(k_c,\omega_c)$ as
\begin{align}
  A^{<}_d(k_c,\omega_c) &= \abs{\phi(\pi-k_c)}^2 \nonumber \\
  &\quad\times \delta(\omega_c + \epsilon_h(\pi-k_c)-U+4t_{\rm hop}+E_b), \label{eq:Exact_Doublon_spectrum}
\end{align}
where the $\pi$-shift in $\epsilon_h$ and $\phi$ comes from the center of mass momentum of the doublon and holon.
The spectrum has a peak at $\omega_c = -\epsilon_h(\pi-k_c)+U-4t_{\rm hop}-E_b$ with the weight $\abs{\phi(\pi-k_c)}^2$, where $-\epsilon_h(\pi-k_c)+U-4t_{\rm hop}$ represents the $\pi$-shifted holon band just below the doublon band.
The formation of the exciton with the binding energy, $-E_b$, opens a gap between the doublon band and the $\pi$-shifted holon band.

Figures~\hyperref[fig:Charge_part_spectra]{8(a)}, \hyperref[fig:Charge_part_spectra]{8(b)}, and \hyperref[fig:Charge_part_spectra]{8(c)} show $A^{<}_d(k_c,\omega_c)$ and $A^{>}_h(-k_c,-\omega_c)$ for $V=0$, $3$, and $5$, respectively. 
$A^{<}_d(k_c,\omega_c)$ is evaluated from Eq.~\eqref{eq:Exact_Doublon_spectrum}, whereas $A^{>}_h(k_c,\omega_c)$, which requires solving the three-body problem of two holons and one doublon, is calculated using exact diagonalization. 
Note that we plot the holon slave-particle spectrum as a function of $-\omega_c$ and $-k_c$ to directly show the holon band, $-\epsilon_h(-k_c)$.
Without an exciton, $A^{<}_d(k_c,\omega_c)$ only shows the signal of the $\pi$-shifted holon band at $k_c=\pi$ due to the sharp peak with the intensity $\abs{\phi(\pi-\pi)}$, see Fig.~\hyperref[fig:Charge_part_spectra]{8(a)}. 
Note that the $\pi$-shifted holon band appears just at the bottom of the doublon band (black dashed line) since the binding energy is zero.
With an exciton, $A^{<}_d(k_c,\omega_c)$ shows the whole $\pi$-shifted holon band due to the broad distribution with the intensity $\abs{\phi(\pi-k_c)}$, see Figs.~\hyperref[fig:Charge_part_spectra]{8(b)} and \hyperref[fig:Charge_part_spectra]{8(c)}.
The $\pi$-shifted holon band is separated from the doublon band by the binding energy $E_b$.

On the other hand, in $A^{>}_h(-k_c,-\omega_c)$ for the case without an exciton, we can only see the holon band (orange dashed line), see Fig.~\hyperref[fig:Charge_part_spectra]{8(a)}.
Additionally, in $A^{>}_h(-k_c,-\omega_c)$ for the case with an exciton, two weak intensity bands emerge above and below the holon band, see Figs.~\hyperref[fig:Charge_part_spectra]{8(b)} and \hyperref[fig:Charge_part_spectra]{8(c)}.
When $V$ increases, the two bands get more separated from the holon band, and become flatter.
The momentum dependence and the position of the flat bands agree well with the exact diagonalization results in Figs.~\hyperref[fig:Spectra_ED]{2(c)} and \hyperref[fig:Spectra_ED]{2(d)}.
The flat bands thus originate from the trion states of two holons and one doublon.
The upper flat band corresponds to the holon-doublon-holon bound state, while the lower one corresponds to the holon-holon-doublon bound state as explained in Sec.~\ref{sec:Spin_annihilation_process}.

\subsection{Spectrum of the spinon slave particle} \label{sec:Spin_part}

\begin{figure}[t!]
  \centering
  \includegraphics[width=0.45\textwidth]{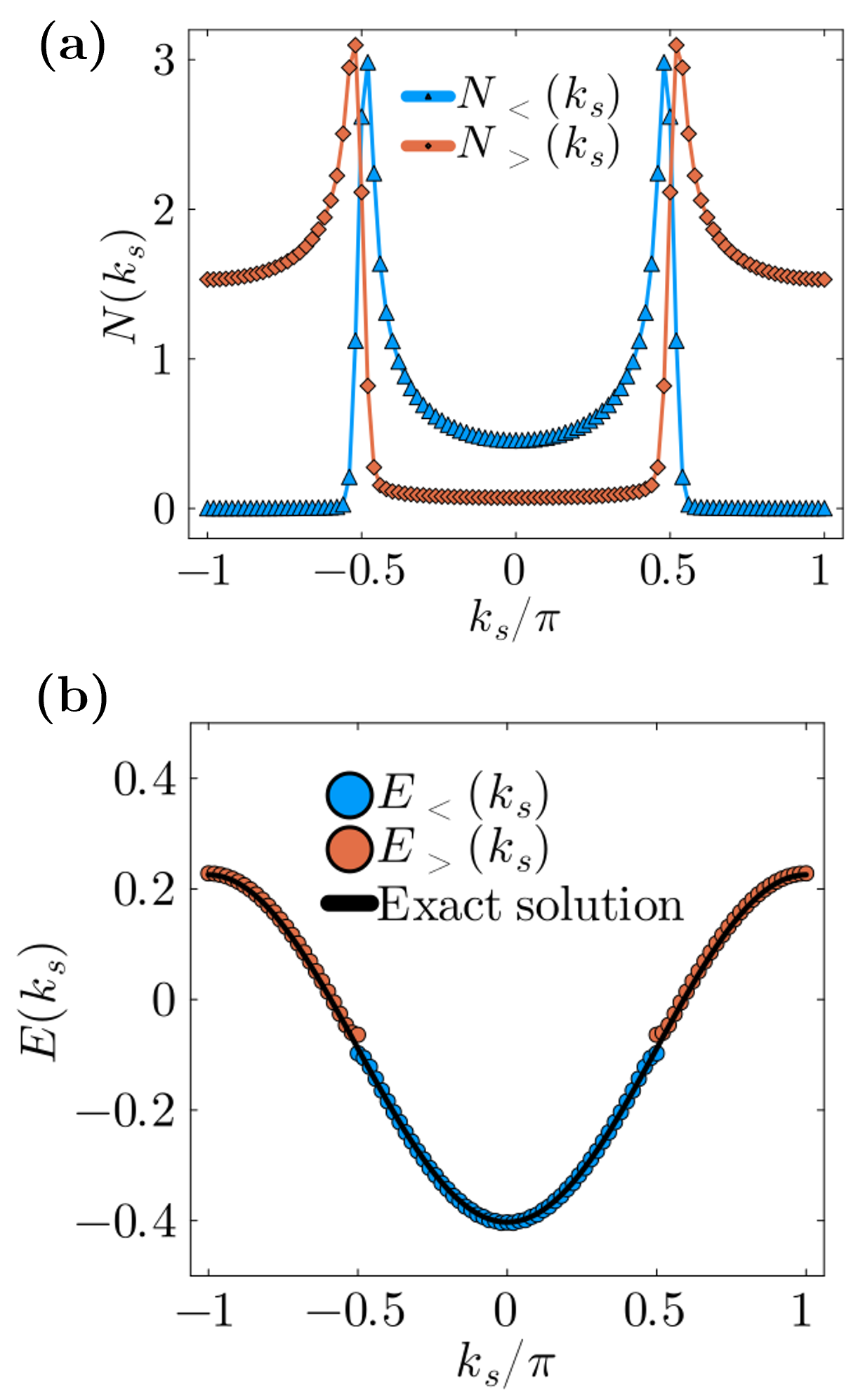}
  \caption{
    (a) The matrix elements $N_{<}(k)$ (blue markers) and $N_{>}(k)$ (orange markers). 
    (b) The energy required to remove the spinon $E_{<}(k)$ (blue dots) and the energy required to add the spinon $E_{>}(k)$ (orange dots).
    The calculation is performed with the DMRG method.
    We also plot the exact spinon dispersion, $-\frac{\pi J}{2}\cos(k) - (\ln2-1/4)J$, obtained from the Bethe-ansatz method (black line). 
    We use $J=4t_{\mathrm{hop}}^2/U=0.2$, $L-2=98$.
    } \label{fig:Spin_part_matele_dispersion}
\end{figure}

In this subsection, we analyze the spectrum of the spinon slave particle by the density matrix renormalization group (DMRG) method.
To construct the spinon spectrum, we calculate the spinon excitation energies and the matrix elements of the spinon creation and annihilation operators. 
We find that the spinon excitation has a gapless dispersion with the spinon Fermi surface at $k_s = \pm \pi/2$.
However, the matrix elements are strongly momentum dependent, which is different from the mean-field picture of free spinons~\cite{Wen_MeanField_2007}.
Note that similar analyses of the spinon spectrum have already been performed based on the exact diagonalization method~\cite{Sorella_Onehole_1992,Penc_Spectral_1995,Penc_Spectral_1997,Penc_Finitetemperature_1997,Kulka_Nature_2025}.
Since we need high resolution of the spinon spectrum to obtain the full ARPES spectra in Sec.~\ref{sec:Convolution_part}, we employ the DMRG method to access large system sizes.

We consider the spin system with $L-2$ sites with the Heisenberg interaction.
In order to calculate the spinon spectrum, we consider the situation of removing (adding) one spin with momentum $k_s$ from (to) the ground state.
Such spinon excitation energy is given by $-\pi J \cos(k_s)/2 - (\ln2-1/4)J$ according to the Bethe-ansatz method \cite{Bethe_Zur_1931,desCloizeaux_SpinWave_1962,Faddeev_What_1981,Talstra_Creation_1997,Kulka_Nature_2025}.
However, it is difficult to calculate the spectral function directly from the Bethe-ansatz solution due to the difficulties in evaluating the matrix elements.
Instead, here we calculate the spectral function numerically with the help of the DMRG method.
If we assume that the contributions to $A^{<}_{Z}(k_s,\omega_s)$ and $A^{>}_{Z}(k_s,\omega_s)$ mostly come from the single spinon excitation, the spectrum of the spinon slave particle can be written as
\begin{align}
  A^{<}_{Z}(k_s,\omega_s) =& -\frac{1}{\pi} \sum_{\sigma} \Im \qty(\frac{ N_{<}(k_s) }{\omega_s + i\eta - E_{<}(k_s)} ), \label{eq:Spinon_occupied_spectrum} \\
  A^{>}_{Z}(k_s,\omega_s) =& -\frac{1}{\pi} \sum_{\sigma} \Im \qty(\frac{ N_{>}(k_s) }{\omega_s + i\eta - E_{>}(k_s)} ), \label{eq:Spinon_unoccupied_spectrum}
\end{align}
where 
\begin{align}
  N_{<}(k_s) =& \bra{s_0} \hat{Z}^{\dagger}_{k_s,\sigma} \hat{Z}_{k_s,\sigma} \ket{s_0}, \\
  N_{>}(k_s) =& \bra{s_0} \hat{Z}_{k_s,\sigma} \hat{Z}^{\dagger}_{k_s,\sigma} \ket{s_0}, \\
  E_{<}(k_s) =& E^{\mathrm{s}}_0 - \frac{\bra{s_0} \hat{Z}^{\dagger}_{k_s,\sigma} \hat{H}_{\mathrm{s}} \hat{Z}_{k_s,\sigma} \ket{s_0}}{N_{<}(k_s)}, \\
  E_{>}(k_s) =& \frac{\bra{s_0} \hat{Z}_{k_s,\sigma} \hat{H}_{\mathrm{s}} \hat{Z}^{\dagger}_{k_s,\sigma} \ket{s_0}}{N_{>}(k_s)} - E^{\mathrm{s}}_0.
\end{align}
Here $\ket{s_0}$ is the ground state of the Heisenberg model $\hat{H}_s$, $N_{<}(k_s)$ ($N_{>}(k_s)$) denotes the matrix element of the spin-removal (addition) operator with momentum $k_s$, and $E_{<}(k_s)$ ($E_{>}(k_s)$) represents the energy difference between the ground state and the one with a spin removed (added) at momentum $k_s$, corresponding to the spinon excitation energy.
Since the spinon-annihilated state $\hat{Z}_{k_s,\sigma}\ket{s_0}$ and the spinon-created state $\hat{Z}^{\dagger}_{k_s,\sigma}\ket{s_0}$ are not normalized, the energy expectation value must be divided by the corresponding norm.

In order to calculate Eqs.~\eqref{eq:Spinon_occupied_spectrum} and \eqref{eq:Spinon_unoccupied_spectrum}, we first calculate $\ket{s_0}$ for the Heisenberg model with the open boundary condition using the DMRG method, for which
we use the ITensor library \cite{Fishman_ITensor_2022}.
The DMRG methods can simulate relatively large systems compared to the exact diagonalization method.
However, in the open boundary condition, we cannot directly apply the Fourier transformation due to the boundary effects.
In order to relax them, we perform the Fourier transform with a Gaussian window function $w(j)=e^{-\alpha j^2}$.
The technical details of the DMRG method are given in Appendix~\ref{sec:Technical_Detail_DMRG}. 

Figure \hyperref[fig:Spin_part_matele_dispersion]{9(a)} shows the obtained matrix elements.
We can see that $N_{<}(k_s)$ ($N_{>}(k_s)$) is strongly suppressed for $\abs{k_s} \geq \pi/2$ ($\abs{k_s} \leq \pi/2$).
Conversely, there are finite matrix elements outside of these regions.
This indicates that the spinon occupies the states only inside of the spinon Fermi surface at $k_s=\pm \pi/2$.
The matrix elements have momentum dependence, and show diverging behavior at $k_s = \pm \pi/2$~\cite{Sorella_Onehole_1992,Penc_Spectral_1995,Penc_Spectral_1997,Penc_Finitetemperature_1997,Kulka_Nature_2025}.
In the mean-field spinon picture, such strong momentum dependence does not appear because the spinon is treated as a free particle (details are given in Appendix~\ref{sec:Mean_field_spinon}).
Therefore, the strong momentum dependence suggests a collective nature of the spinon excitation beyond the mean-field picture.

Figure \hyperref[fig:Spin_part_matele_dispersion]{9(b)} shows the energy difference, $E_{<}(k_s)$ ($E_{>}(k_s)$),
which is plotted only for $\abs{k_s} \leq \pi/2$ ($\abs{k_s} \geq \pi/2$) since $N_{<}(k_s)$ ($N_{>}(k_s)$) is almost zero outside of these regions.
We also plot the exact spinon dispersion, $-\pi J\cos(k_s)/2 - (\ln2-1/4)J$.
We can see that the energy differences well agree with the exact spinon dispersion, which justifies the above-mentioned assumption of the single spinon contribution.

\begin{figure}[t!]
  \centering
  \includegraphics[width=0.45\textwidth]{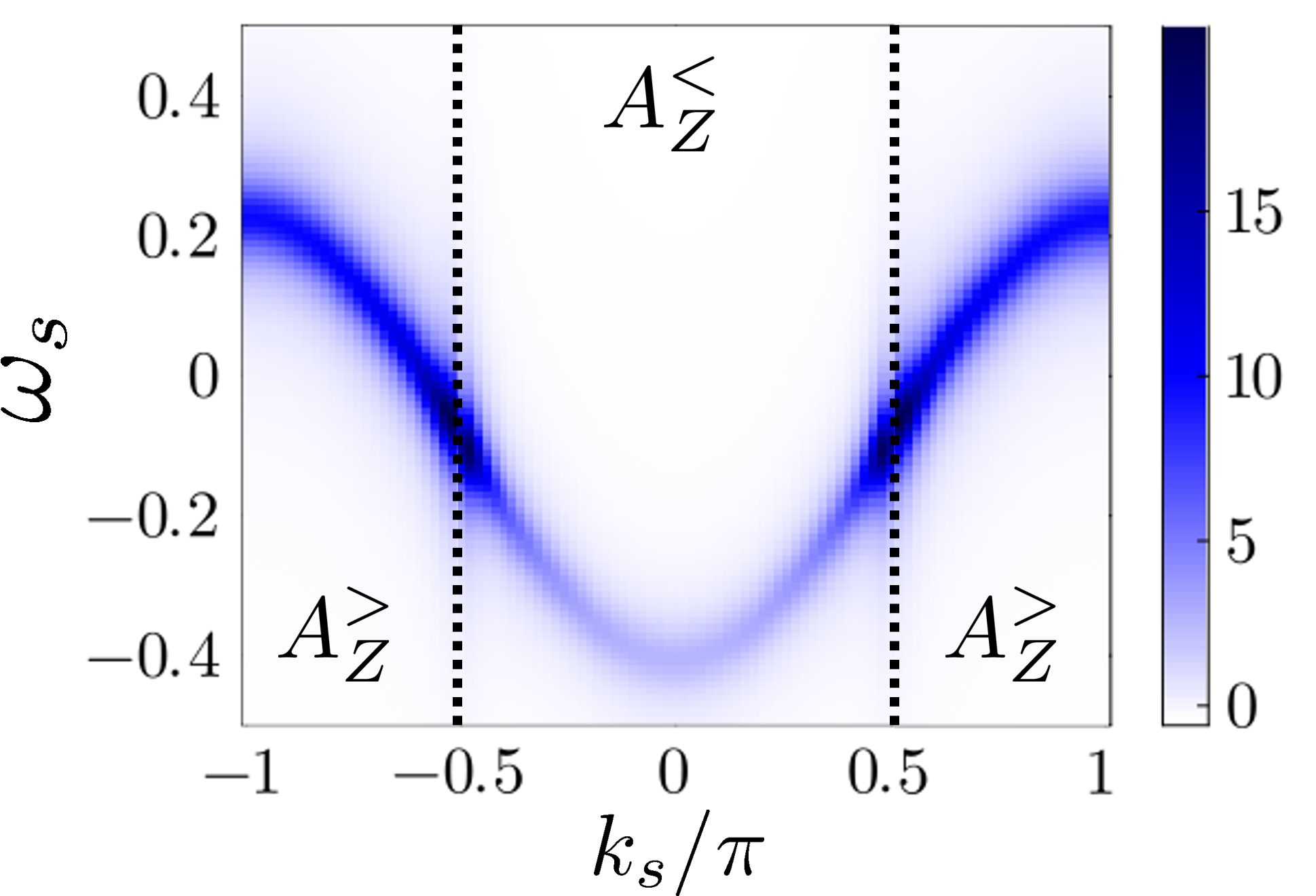}
  \caption{
    Occupied (unoccupied) spectrum of the spinon slave particle $A^{<}_Z(k_s,\omega_s)$ ($A^{>}_Z(k_s,\omega_s)$).
    We show the occupied spectrum for $\abs{k_s} \leq \pi/2$ and the unoccupied spectrum for $\abs{k_s} \geq \pi/2$.
    The calculation is based on Eqs.~\eqref{eq:Spinon_occupied_spectrum}, \eqref{eq:Spinon_unoccupied_spectrum} and the results shown in Fig.~\ref{fig:Spin_part_matele_dispersion}.
    Here we use $J=4t_{\mathrm{hop}}^2/U=0.2$, $L-2=98$.
    } \label{fig:Spin_part_spectrum}
\end{figure}

Figure~\ref{fig:Spin_part_spectrum} shows the spectrum of the spinon slave particle based on Eqs.\eqref{eq:Spinon_occupied_spectrum} and \eqref{eq:Spinon_unoccupied_spectrum}.
We show $A^{<}_Z(k_s,\omega_s)$ only for $\abs{k_s} \leq \pi/2$ and $A^{>}_Z(k_s,\omega_s)$ only for $\abs{k_s} \geq \pi/2$.
Note that the spectral weights are almost zero outside these regimes.
The signals appear along the gapless spinon dispersion.
The spectrum of the spinon slave particle has a high intensity around $k_s=\pm \pi/2$, which reflects the diverging behavior of the matrix elements.

\subsection{Convolution of the slave-particle spectra} \label{sec:Convolution_part}

\begin{figure*}[t!]
  \centering
  \includegraphics[width=0.9\textwidth]{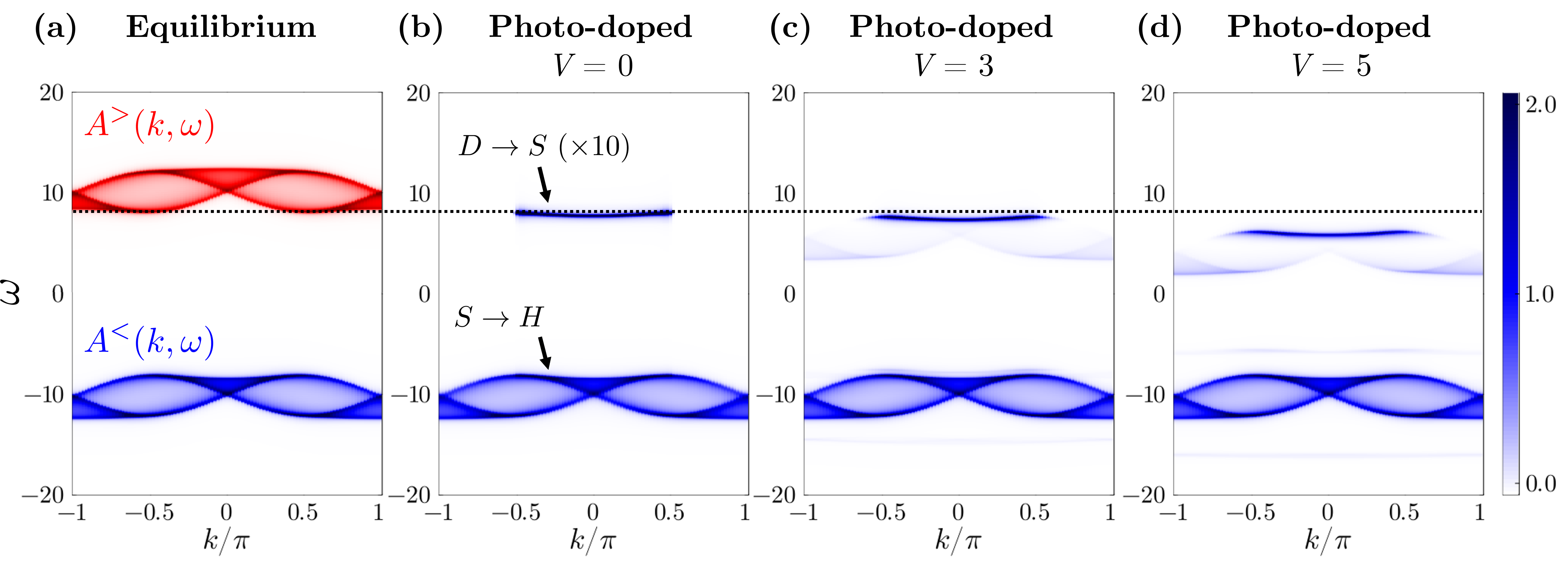}
  \caption{
    (a) Occupied spectral function $A^{<}(k,\omega)$ (blue color map) and unoccupied spectral function $A^{>}(k,\omega)$ (red) for equilibrium Mott insulators ($N_d=N_h=0$) within the slave particle approach.
    The black dashed line shows the band edge of the UHB.
    (b), (c), (d) Occupied spectrum $A^{<}(k,\omega) = A^{<}_{\mathrm{D\to S}}(k,\omega) + A^{<}_{\mathrm{S\to H}}(k,\omega)$ for photodoped Mott insulators ($N_d=N_h=1$) with (b) $V=0$, (c) $V=3$, and (d) $V=5$ within the slave particle approach.
    The calculations are based on the convolution of the slave particle spectra shown in Figs.~\ref{fig:Charge_part_spectra} and \ref{fig:Spin_part_spectrum}.
    For the sake of visibility, $A^{<}_{\mathrm{D\to S}}(k,\omega)$ is multiplied by 10.
    Here we use $U=20$, and $L=100$.
    } \label{fig:Convolution}
\end{figure*}

Finally, we calculate the photoemission spectra by taking the convolution of the slave particle spectra based on Eqs.~\eqref{eq:Spectral_decomposition1} and \eqref{eq:Spectral_decomposition2}.
The results are shown in Fig.~\ref{fig:Convolution}, where we plot the occupied and unoccupied spectra for an equilibrium Mott insulator [Fig.~\ref{fig:Convolution}(a)] and occupied spectra for photodoped Mott insulators with (b) $V=0$, (c) $V=3$, and (d) $V=5$.
We find that the slave-particle analysis successfully reproduces the following characteristics observed in the photoemission spectra obtained by ED:
\begin{enumerate}
    \item When a doublon and a holon are not bound, a dispersive in-gap signal emerges just below the UHB, see $A^{<}_{\mathrm{D\to S}}(k,\omega)$ in Fig.~\hyperref[fig:Convolution]{11(b)}.
    \item When a Mott-Hubbard exciton is formed, a replica LHB signal appears below the UHB.
    The intensity of this signal becomes more broadly distributed as $V$ is increased, see $A^{<}_{\mathrm{D\to S}}(k,\omega)$ in Figs.~\hyperref[fig:Convolution]{11(c)} and \hyperref[fig:Convolution]{11(d)}.
    \item When a Mott-Hubbard exciton is formed, two flat bands appear above and below the LHB, see $A^{<}_{\mathrm{S\to H}}(k,\omega)$ in Figs.~\hyperref[fig:Convolution]{11(c)} and \hyperref[fig:Convolution]{11(d)}.
\end{enumerate}
We now discuss the physical origin of these features within the slave-particle picture.

We first discuss the spectral feature (i).
Based on Figs.~\hyperref[fig:Charge_part_spectra]{8(a)} and \ref{fig:Spin_part_spectrum}, we can explain the origin of the dispersive in-gap signal as follows.
In the absence of the doublon-holon binding, the doublon has a well-defined momentum $k_c = \pi$ and energy at the bottom of the doublon band.
On the other hand, the created spinon takes various momentum states ($\abs{k_s} \gtrapprox \pi/2$) above the spinon Fermi surface.
Due to the momentum and energy conservations ($k = k_c - k_s$ and $\omega = \omega_c - \omega_s$), the spinon dispersion appears for $\abs{k} \lessapprox \pi/2$ just below the UHB.

Second, we discuss the spectral feature (ii).
Based on Figs.~\hyperref[fig:Charge_part_spectra]{8(b)}, \hyperref[fig:Charge_part_spectra]{8(c)} and \ref{fig:Spin_part_spectrum}, the origin of the replica LHB below the UHB can be understood as follows.
When a Mott-Hubbard exciton is formed, the doublon can occupy various momentum states due to its localized nature in real space, as can be seen from Fig.~\hyperref[fig:Charge_part_wave_Eb]{7(a)}.
Thus, the LHB-like structure of the in-gap signal can be understood by moving the doublon occupied spectra (Figs.~\hyperref[fig:Charge_part_spectra]{8(b,c)}) along the spinon unoccupied spectrum (Fig.~\ref{fig:Spin_part_spectrum}).
Note that the structure of the LHB in equilibrium can also be understood in a similar way~\cite{Bohrdt_Angleresolved_2018}.
The binding energy of the Mott-Hubbard exciton lowers the energy cost of photoemission from the doublon, thereby shifting the photoemission signal downward by the binding energy.
We also comment on the quantitative differences between the exact diagonalization results and the slave-particle convolution results.
The replica LHB calculated by the slave-particle method shows a finite intensity around $k=\pi$, whereas the exact diagonalization result does not.
This discrepancy may originate from the approximations in Eqs.~\eqref{eq:H_spin_ex_approx} and \eqref{eq:Slave_particle_representation_momentum}, which neglect the correlation between the charge and spin degrees of freedom in the squeezed space.

Third, we discuss the spectral feature (iii).
The holon slave-particle spectra in Figs.~\hyperref[fig:Charge_part_spectra]{8(b)} and \hyperref[fig:Charge_part_spectra]{8(c)} 
successfully reproduce the position and intensity of the two flat bands observed in the exact diagonalization results.
On the other hand, the convolution with the spinon spectrum slightly distorts the flatness and momentum-dependent intensity of the flat bands, which differs from the exact diagonalization results.
This discrepancy may originate from the approximations in Eqs.~\eqref{eq:H_spin_ex_approx} and \eqref{eq:Slave_particle_representation_momentum}.
According to the semiclassical picture of the trion excitation explained in Sec.~\ref{sec:Spin_annihilation_process}, trion creation does not induce spinon excitation in the squeezed space.
Therefore, the unconditional convolution of the charge and spin spectra may not be valid for the trion excitation.
How to improve the approximation is left for future work.

\section{Conclusion and outlook} \label{sec:Conclusions}

In this paper, we have studied the momentum-resolved photoemission spectra of photodoped one-dimensional Mott insulators, both with and without exciton formation, using the steady-state formulation. 
Reflecting strong correlation effects, photodoped Mott insulators exhibit distinct characteristic spectral features in both cases. 
In the absence of exciton formation, a dispersive in-gap signal appears just below the UHB. When an exciton is formed, a complex in-gap signal resembling the LHB appears, together with two flat bands located above and below the LHB.

We have clarified the origin of these spectral features from two complementary viewpoints: the real-time dynamics of elementary excitations triggered by photoemission and the spectral decomposition within the slave-particle representation. 
From the viewpoint of elementary excitation dynamics, in the absence of exciton formation, the dispersive in-gap signal originates from spinon propagation created by photoemission from the doublon. 
When an exciton is present, photoemission from the doublon induces both spinon and holon propagation, which gives rise to the replica of the LHB. 
Furthermore, photoemission from a singlon near the exciton creates two types of trions, leading to the two flat bands. 
Within the slave-particle description, the dispersive in-gap signal arises from spinon excitations above the spinon Fermi surface. 
The replica of the LHB reflects the broad momentum distribution of the Mott–Hubbard exciton wavefunction, while the two flat bands originate from massive trion excitations created by singlon annihilation. 
These two approaches provide a comprehensive understanding of the spectral features.

Our results suggest that time-resolved photoemission spectroscopy in strongly correlated materials can serve as a useful experimental probe of both the binding properties of Mott-Hubbard excitons and the underlying magnetic correlations. 
To further establish and broaden this perspective, it will be important to investigate other types of Mott-Hubbard excitons in a wider range of correlated electron systems. 
In this regard, the steady-state formulation employed here provides a promising framework for systematic and comprehensive studies beyond the computational limits of fully real-time simulations.

An interesting direction is to extend our analysis to higher-dimensional systems.
In higher dimensions, the spin degree of freedom and the charge degree of freedom are coupled to each other.
Therefore, the photoemission spectra may not be understood as simple convolutions of the spin and charge spectra, and an interpretation based on the slave-particle picture of spin-charge separation may not be directly applicable.
Clarifying how spin-charge coupling affects the photoemission spectra, and how the spectral features differ from the one-dimensional case, is an important future direction.

\acknowledgments
We are grateful to K. M. Dani, S. Imai, T. Kaneko, H. Katsura, T. Morimoto, K. Takasan, S. Takayoshi, and P. Werner for helpful discussions.
This work is supported by JSPS KAKENHI (Grant Nos.~JP21H05017, JP24H00191, JP25H01246, JP25H01251, No. JP25K07235, No. JP26H01281, and No. JP26K00646) and JST FOREST (Grant No.~JPMJFR2131).
T. N. is supported by WINGS-MERIT of the University of Tokyo and JST SPRING (Grant No.~JPMJSP2108).

\appendix

\section{Photoemission spectra for \texorpdfstring{$U=10$}{TEXT}} \label{sec:Spectrum_U10}

\begin{figure*}[t]
  \centering
  \includegraphics[width=0.95\textwidth]{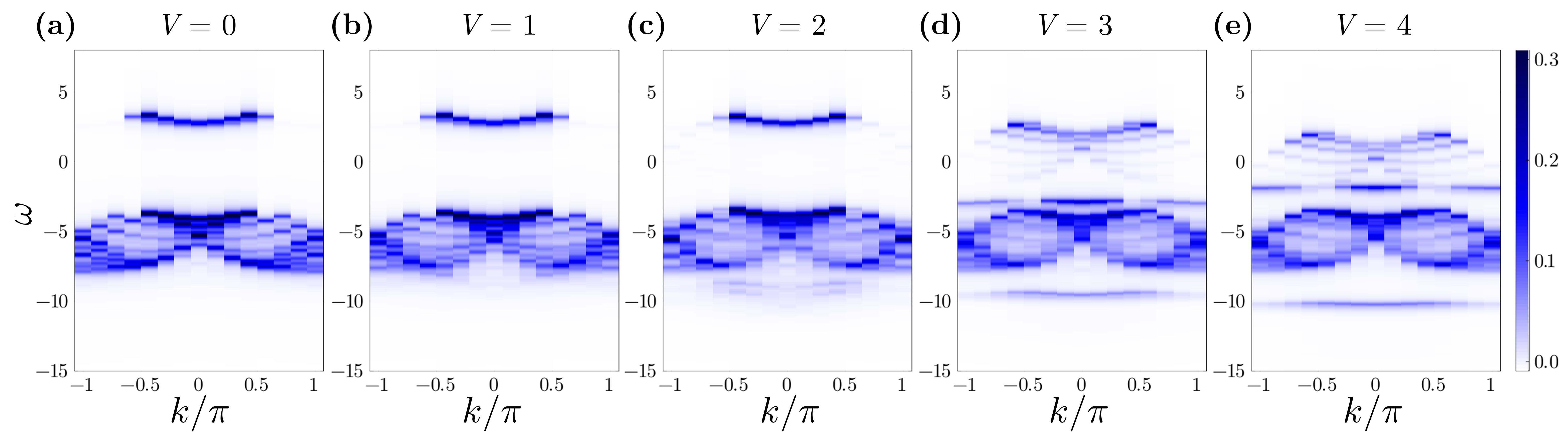}
  \caption{
    Photoemission spectra $A^{<}(k,\omega)$ of photodoped Mott insulators with $N_d=N_h=1$, for (a) $V=0$, (b) $V=1$, (c) $V=2$, (d) $V=3$, and (e) $V=4$.
    Here we use $U=10$, $L=14$, and $\eta=0.15$.
    } \label{fig:Spectra_U10}
\end{figure*}

In this Appendix, we show the photoemission spectra for $U=10$.
In the main text, we have shown the results for $U=20$ to clearly see the in-gap spectrum and the two flat bands.
However, a typical value of the on-site interaction in realistic Mott insulators lies about $U\sim10$.
Here we show the results for $U=10$ to confirm that the behavior of the spectra is qualitatively the same as in the case of $U=20$.

Figures \hyperref[fig:Spectra_U10]{12(a)-(e)} show $A^{<}(k,\omega)$ for $V=0,1,2,3,4$, respectively. 
We can see that the shape and position of the spinon dispersion remain almost the same as long as $V \lesssim 2$, where no exciton is formed.
The copy of the LHB and the two flat bands appear when $V = 3$ and $4$, where the exciton is formed. 
The two bands are not perfectly flat at $V=3$, but become flatter at $V=4$.
Our result at $U=10$ and $V=3$ corresponds to that of the recent pump-probe simulation shown in Fig.~5(g) of Ref.~\cite{Sugimoto_Pumpprobe_2023}, and they agree well with each other.

Compared to the $U=20$ case, a clear spectral separation between the replica LHB band and the upper flat band is not observed due to the smaller Mott gap.
However, the overall spectral trend is consistent with the $U=20$ case, which supports the experimental relevance of our analysis.

\section{Momentum dependence of the spectral weight of trions} \label{sec:Trion_excitation_energy}

\begin{figure}[t!]
  \centering
  \includegraphics[width=0.45\textwidth]{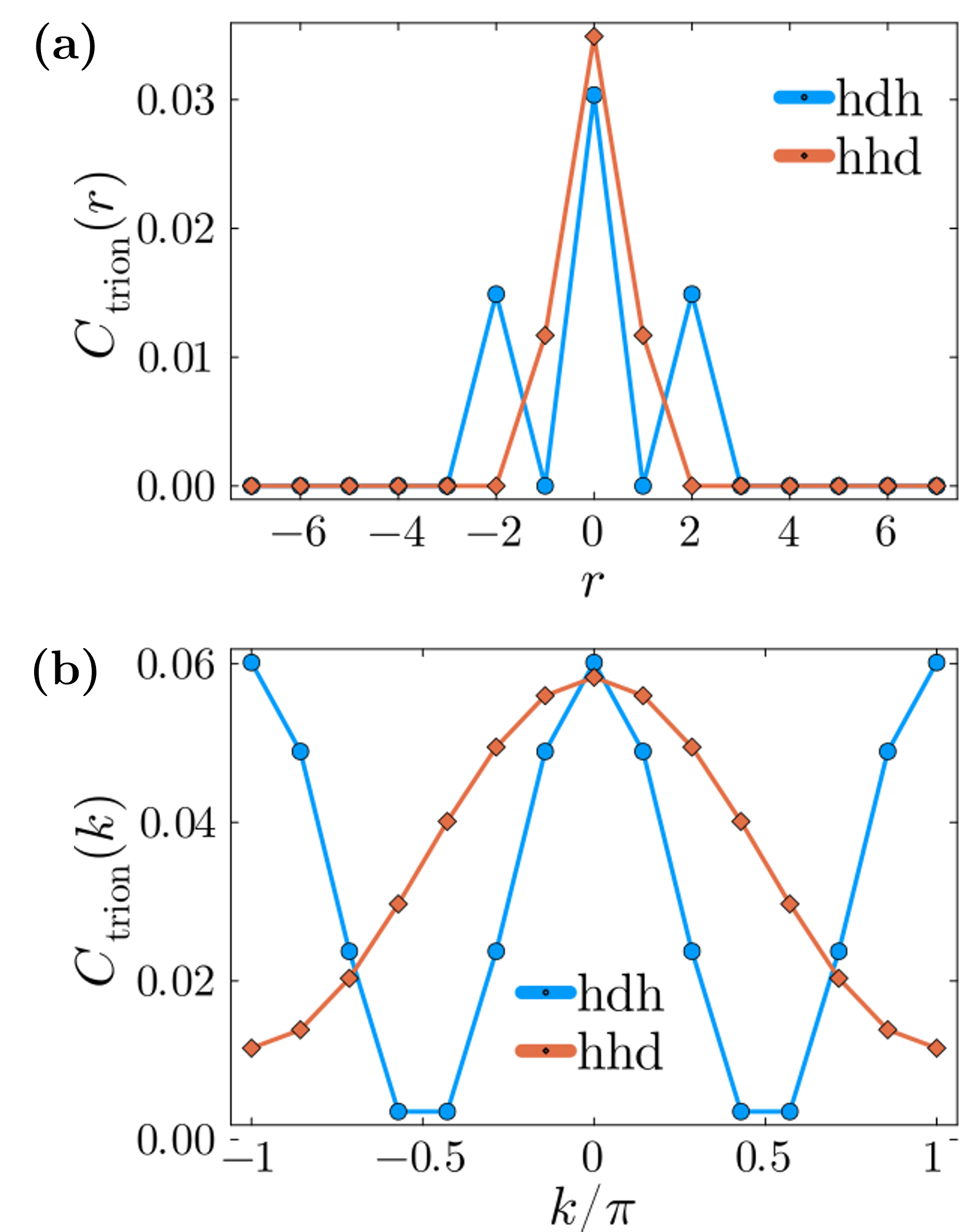}
  \caption{
    (a) Correlation function $C_{\mathrm{trion}}(r)$ of the hdh-trion (blue line) and the hhd-trion (orange).
    (b) Fourier transformed correlation function $C_{\mathrm{trion}}(k)$ of the hdh-trion (blue line) and the hhd-trion (orange). 
    Here we use $U=20$, $V=5$, and $L=14$.
    } \label{fig:Trion_momentum}
\end{figure}

In this Appendix, we explain the momentum dependence of the photoemission spectrum from two trions shown in Fig.~\hyperref[fig:Spectra_ED]{2(d)}.
If we focus on the $\alpha$-trion state ($\alpha$ refers to ``hdh'' or ``hhd'') $\ket{\psi^{\alpha-\mathrm{trion}}_m}$ and assume that the trion energy $E_{\alpha-\mathrm{trion}}$ is independent of momentum, the photoemission spectrum from the trion state is given by
\begin{align}
  A^{<}_{\alpha-\mathrm{trion}}(k,\omega) \approx& \sum_{\sigma} \sum_m \abs{\bra{\psi^{\alpha-\mathrm{trion}}_m} \hat{c}_{k,\sigma} \ket{\psi_0}}^2 \nonumber \\
  &\delta(\omega - E_{\alpha-\mathrm{trion}}), \\
  =& \sum_{\sigma,i,j} e^{ik(i-j)} \bra{\psi_0} \hat{c}_{i,\sigma}^{\dagger} P_{\alpha-\mathrm{trion}} \hat{c}_{j,\sigma} \ket{\psi_0} \nonumber \\
  &\delta(\omega - E_{\alpha-\mathrm{trion}}),
\end{align}
where $P_{\alpha-\mathrm{trion}} = \sum_m \ket{\psi^{\alpha-\mathrm{trion}}_m}\bra{\psi^{\alpha-\mathrm{trion}}_m}$ is the projection operator onto the $\alpha$-trion states.
Therefore, momentum dependence is determined by the correlation function,
\begin{align}
  C_{\alpha-\mathrm{trion}}(r) = \sum_{\sigma} \bra{\psi_0} \hat{c}_{j+r,\sigma}^{\dagger} P_{\alpha-\mathrm{trion}} \hat{c}_{j,\sigma} \ket{\psi_0}.
\end{align}
Due to the translational symmetry in the periodic boundary condition, $C_{\alpha-\mathrm{trion}}(r)$ is independent of $j$.

Figure~\hyperref[fig:Trion_momentum]{13(a)} shows $C_{\alpha-\mathrm{trion}}(r)$ for the hdh-trion and hhd-trion for $V=5$.
Due to the trion structure, $C_{\mathrm{hdh-trion}}(r)$ ($C_{\mathrm{hhd-trion}}(r)$) is nonzero only for $r=0,\pm2$ ($r=0,\pm1$).
Therefore, the Fourier transformed correlation function $C_{\mathrm{hdh-trion}}(k)$ ($C_{\mathrm{hhd-trion}}(k)$) has weak intensity around $k=\pm \pi/2$ ($k=\pm \pi$), as shown in Fig.~\hyperref[fig:Trion_momentum]{13(b)}.

\section{Derivation of the slave-particle representation in the squeezed space} \label{sec:Slave_particle_squeezed_space}

In this Appendix, we derive the Hamiltonian in the squeezed space representation using the slave-particle method.
First, we rewrite the Hamiltonian \eqref{eq:Effective_Hamiltonian} using the slave-particle representation in the original lattice.
Using the slave-particle representation in Eq.~\eqref{eq:Slave_particle_representation} and the local constraint in Eq.~\eqref{eq:Local_constraint}, we can translate the original electron operators as
\begin{align}
  \hat{c}_{j,\sigma}\hat{n}_{j,\bar{\sigma}} &= (-1)^\sigma \hat{d}_j \hat{f}^{\dagger}_{j,\bar{\sigma}}, \\
  \hat{c}_{j,\sigma}(1-\hat{n}_{j,\bar{\sigma}}) &= \hat{h}^{\dagger}_j \hat{f}_{j,\sigma}, \\
  (\hat{n}_{j,\uparrow} - 1/2)(\hat{n}_{j,\downarrow} - 1/2) &= \frac{1}{2}\qty(\hat{n}^{\mathrm{d}}_j + \hat{n}^{\mathrm{h}}_j - \frac{1}{2}), \\
  \hat{\eta}^+_j &= (-1)^j \hat{d}^{\dagger}_j \hat{h}_j, \\
  \hat{\eta}^z_j &= \frac{1}{2}(\hat{n}^{\mathrm{d}}_j - \hat{n}^{\mathrm{h}}_j), \\ 
  \hat{\vb{s}}_j &= \frac{1}{2} \sum_{\alpha,\beta} \hat{f}^{\dagger}_{j,\alpha} \sigma_{\alpha,\beta} \hat{f}_{j,\beta}.
\end{align}
Then, the effective Hamiltonian is rewritten as Eqs.~\eqref{eq:H_U}, \eqref{eq:H_V_dh_ex}, and
\begin{align}
  \hat{H}_{\mathrm{kin,d}} = & t_{\mathrm{hop}} \sum_{j,\sigma} \hat{f}^{\dagger}_{j+1,\bar{\sigma}} \hat{d}^{\dagger}_{j} \hat{d}_{j+1} \hat{f}_{j,\bar{\sigma}} + \mathrm{h.c.}, \\
  \hat{H}_{\mathrm{kin,h}} = & -t_{\mathrm{hop}} \sum_{j,\sigma} \hat{f}^{\dagger}_{j+1,\sigma} \hat{h}^{\dagger}_{j} \hat{h}_{j+1} \hat{f}_{j,\sigma} + \mathrm{h.c.}, \\
  \hat{H}_{\mathrm{spin,ex}} =& J \sum_{j} \hat{\vb{s}}_j \cdot \hat{\vb{s}}_{j+1}. 
\end{align}
In the doublon hopping term, we use the anticommutation relation of fermion operators.
Therefore, the doublon hopping strength and the holon hopping strength have the opposite sign.

Second, we translate the Hamiltonian into the squeezed-space representation using $\hat{f}_{j,\sigma} = \hat{Z}_{l(j),\sigma}$.
Considering the doublon (holon) hopping process $\hat{d}^{\dagger}_{j} \hat{d}_{j+1}$ ($\hat{h}^{\dagger}_{j} \hat{h}_{j+1}$), the relation $l(j) = l(j+1)$ holds because $j$ or $j+1$ is always occupied by the doublon (holon).
This means that the spin configuration in the squeezed space does not change during the doublon (holon) hopping process.
Then, the kinetic terms are rewritten as Eqs.~\eqref{eq:H_kin_d} and \eqref{eq:H_kin_h} in the main text.
The spin-spin interaction term acts only when both $j$ and $j+1$ are occupied by singlons because of the local constraint in Eq.~\eqref{eq:Local_constraint}.
Instead of considering the local constraint, we only consider whether spins $l$ and $l+1$ occupy nearest-neighbor sites in the original lattice or not.
This can be expressed as $\delta_{j(l)+1, j(l+1)}$, where $j(l)= \mathrm{max}\{m | l(m) = l\}$ is the original lattice site index as a function of the squeezed space site index.
Then, the spin-spin interaction term is rewritten as Eq.~\eqref{eq:H_spin_ex} in the main text.

\section{Derivation of the spectral decomposition in the slave-particle method} \label{sec:Spectra_decomposition}

In this Appendix, we derive the approximate decomposition formulas of the spectral function Eqs.~\eqref{eq:Spectral_decomposition1} and \eqref{eq:Spectral_decomposition2} in the slave-particle approach.
The Lehmann representation of the doublon annihilation (holon creation) spectrum is given by
\begin{align}
  A^{<}_{\mathrm{D\to S}}(k,\omega) =& \sum_{n,\sigma} \delta(\omega + E_{n0}) \abs{\bra{\psi_n}\hat{c}_{k,\sigma,\mathrm{D\to S}} \ket{\psi_0}}^2 \label{eq:Lehmann_doublon}, \\
  A^{<}_{\mathrm{S\to H}}(k,\omega) =& \sum_{n,\sigma} \delta(\omega + E_{n0}) \abs{\bra{\psi_n}\hat{c}_{k,\sigma,\mathrm{S\to H}} \ket{\psi_0}}^2 \label{eq:Lehmann_spin},
\end{align}
where $E_{n0} = E_n - E_0$ is the energy difference between the $n$-th excited state and the original photodoped state, and $\ket{\psi_n}$ is the $n$-th excited state with energy $E_n$.

First, we derive the approximate expression of the electron operators  Eq.~\eqref{eq:Slave_particle_representation_momentum}.
In the slave-particle representation, the electron annihilation operator of the process $D\to S$ ($S\to H$) is expressed as $\hat{c}_{j,\sigma,\mathrm{D\to S}} = (-1)^{\sigma} \hat{d}_j \hat{Z}^{\dagger}_{l(j),\bar{\sigma}}$ ($\hat{c}_{j,\sigma,\mathrm{S\to H}} = \hat{h}^{\dagger}_j \hat{Z}_{l(j),\sigma}$).
Here $l(j)= j - \sum_{m<j}(\hat{d}^{\dagger}_m \hat{d}_m + \hat{h}^{\dagger}_m \hat{h}_m)$ is the relabeled site index in the squeezed space.
If there are only one doublon and one holon in the system, the expectation value of the doublon and holon number at site $m<j$ is about $2j/L$.
Therefore, we can approximate $l(j) \approx j(1-\frac{2}{L})$.
The Fourier transform of the squeezed space operator is 
\begin{align}
  \hat{Z}_{l(j),\sigma} \approx& \frac{1}{\sqrt{L}} \sum_{p} e^{-i p j(1-\frac{2}{L})} \hat{Z}_{p,\sigma} \nonumber \\
  = & \frac{1}{\sqrt{L}} \sum_{k_s} e^{-i k_s j} \hat{Z}_{k_s,\sigma}, \label{eq:Fourier_transform_squeezed_space}
\end{align}
where $p=\frac{2\pi}{L-2}m$ ($m \in \mathbb{Z}$) is the momentum in the squeezed space.
We now define $k_s=p\qty(1-\frac{2}{L}) = \frac{2\pi}{L}m$ as the momentum in the original space.
Using Eq.\eqref{eq:Fourier_transform_squeezed_space}, the Fourier transforms of the electron annihilation operators for the processes $D\to S$ and $S\to H$ are expressed as
\begin{align}
  c_{k,\sigma,\mathrm{D\to S}} =& (-1)^{\sigma} \frac{1}{\sqrt{L}} \sum_{k_s} \hat{d}_{k_s+k} \hat{Z}^{\dagger}_{k_s,\bar{\sigma}}, \\
  c_{k,\sigma,\mathrm{S\to H}} =& \frac{1}{\sqrt{L}} \sum_{k_s} \hat{h}^{\dagger}_{k_s-k} \hat{Z}_{k_s,\sigma},
\end{align}
which are summarized into Eq.\eqref{eq:Slave_particle_representation_momentum}.

Furthermore, we assume that the spin-charge coupling in the Hamiltonian $\hat{H}_{\rm eff}$ in the squeezed-space representation can be decoupled into a charge part $\hat{H}_c$, and a spin part $\hat{H}_s$. 
Namely, any eigenstate can be expressed as a product state of the charge part $\ket{c_m}$ and the spin part $\ket{s_l}$,
$\ket{\psi_n} = \ket{c_m} \otimes \ket{s_l}$.
Note that $n$ can now be regarded as a composite index for $(m,l)$, and $\ket{s_l}$ is defined in the squeezed space.
Substituting these into the Lehmann representation in Eqs.~\eqref{eq:Lehmann_doublon} and \eqref{eq:Lehmann_spin}, we obtain
\begin{align}
  A^{<}_{\mathrm{D\to S}}(k,\omega) =& \frac{1}{L}\sum_{n,\sigma}\sum_{k_s} \delta(\omega + E_{n0}) \nonumber \\
  &\times \abs{\bra{c_m}\hat{d}_{k_s+k}\ket{c_0}}^2 \abs{\bra{s_l}\hat{Z}^{\dagger}_{k_s,\bar{\sigma}}\ket{s_0}}^2, \\
  A^{<}_{\mathrm{S\to H}}(k,\omega) =& \frac{1}{L}\sum_{n,\sigma}\sum_{k_s} \delta(\omega + E_{n0}) \nonumber \\
  &\times \abs{\bra{c_m}\hat{h}^{\dagger}_{k_s-k}\ket{c_0}}^2 \abs{\bra{s_l}\hat{Z}_{k_s,\sigma}\ket{s_0}}^2, 
\end{align}
where we use the conservation law for charge and spin momentum.
We can rewrite the energy difference as $E_{n0} = E^{c}_{m0} + E^{s}_{l0}$, where $E^{c}_{m0} = E^{c}_m - E^{c}_0$ is the energy difference in the charge part and $E^{s}_{l0} = E^{s}_l - E^{s}_0$ is the energy difference in the spin part.
We insert the identity $\int d\omega_c d\omega_s \delta(\omega_c+E^c_{m0}) \delta(\omega_s-E^s_{l0}) \sum_{k_c} \delta_{k_c,k_s+k}$ in the doublon annihilation spectrum and $\int d\omega_c d\omega_s \delta(\omega_c-E^c_{m0}) \delta(\omega_s+E^s_{l0}) \sum_{k_c} \delta_{k_c,k_s-k}$ in the holon creation spectrum to obtain
\begin{align}
  A^{<}_{\mathrm{D\to S}}(k,\omega) =& \frac{1}{L} \sum_{k_s,k_c,\sigma}\int d\omega_c d\omega_s \nonumber \\
  &\times \delta_{k,k_c-k_s} \delta(\omega - (\omega_c - \omega_s)) \nonumber \\
  &\times \sum_m \delta(\omega_c + E^c_{m0}) \abs{\bra{c_m}\hat{d}_{k_c}\ket{c_0}}^2 \nonumber \\
  &\times \sum_{l,\sigma} \delta(\omega_s - E^s_{l0}) \abs{\bra{s_l}\hat{Z}^{\dagger}_{k_s,\bar{\sigma}}\ket{s_0}}^2, \\
  A^{<}_{\mathrm{S\to H}}(k,\omega) =& \frac{1}{L}\sum_{k_s,k_c,\sigma}\int d\omega_c d\omega_s \nonumber \\
  &\times \delta_{k,k_s-k_c} \delta(\omega - (\omega_s - \omega_c)) \nonumber \\
  &\times \sum_m \delta(\omega_c - E^c_{m0}) \abs{\bra{c_m}\hat{h}^{\dagger}_{k_c}\ket{c_0}}^2 \nonumber \\
  &\times \sum_{l,\sigma} \delta(\omega_s + E^s_{l0}) \abs{\bra{s_l}\hat{Z}_{k_s,\sigma}\ket{s_0}}^2.
\end{align}
These are nothing but the spectral decomposition formulas \eqref{eq:Spectral_decomposition1} and \eqref{eq:Spectral_decomposition2} in the main text.

\section{Exact solution of the doublon-holon model} \label{sec:Exact_dh_solution}

In this Appendix, we derive the exact solution of the doublon-holon model $\hat{H}_c$ consisting of Eqs.~\eqref{eq:H_kin_d}, \eqref{eq:H_kin_h}, \eqref{eq:H_U}, and \eqref{eq:H_V_dh_ex} for the sector with one doublon and one holon. 
We consider the wavefunction in the form of $\ket{\psi} = \sum_{l,m} \Psi(l,m) \hat{d}^{\dagger}_{l} \hat{h}^{\dagger}_{m} \ket{0_c}$.
The wavefunction satisfies $\Psi(l,l)=0$ due to the hard-core boson condition.
The wavefunction has the particle-hole symmetry, where the Hamiltonian is invariant under the particle-hole transformation, $\hat{h}_m \rightarrow (-1)^m \hat{d}_m$ and $\hat{d}_m \rightarrow (-1)^m \hat{h}_m$.
Therefore, the wavefunction should satisfy the relation $\Psi(l,m) = \pm (-1)^{l+m} \Psi(m,l)$, where $+$ ($-$) corresponds to the even (odd) particle-hole parity.
Here, we focus on the even particle-hole parity state
\begin{align}
  \Psi(l,m) = (-1)^{l+m} \Psi(m,l). \label{eq:PH_symmetry}
\end{align}
because the ground state belongs to the even particle-hole symmetry when the system size $L$ is even.

In the following, we find the wavefunction $\Psi(l,m)$ for $l<m$ because that for $l>m$ can be automatically obtained by Eq.~\eqref{eq:PH_symmetry}.
The eigenvalue equation for $m \neq l+1$ is given by
\begin{align}
  E' \Psi(l,m) = t_{\mathrm{hop}} (& \Psi(l+1,m) + \Psi(l-1,m) \nonumber \\
   - & \Psi(l,m+1) - \Psi(l,m-1)), \label{eq:eigen_neq}
\end{align}
where $E' = E - U$ and $E$ is the eigenenergy.
For $m = l+1$, the eigenvalue equation is given by
\begin{align}
  E' \Psi(l,l+1) =& t_{\mathrm{hop}} ( \Psi(l-1,l+1) - \Psi(l,l+2)) \nonumber \\
   &-  (V-\frac{J}{4})\Psi(l,l+1) + \frac{J}{2}\Psi(l+1,l) \nonumber \\
   =&t_{\mathrm{hop}} ( \Psi(l-1,l+1) - \Psi(l,l+2)) \nonumber \\
   &-  V_{\mathrm{eff}}\Psi(l,l+1)
   \label{eq:eigen_eq}
\end{align}
where $V_{\mathrm{eff}} = V + \frac{J}{4}$ is the effective doublon-holon attractive interaction.
Here we use the particle-hole symmetry condition \eqref{eq:PH_symmetry}.
Equations \eqref{eq:eigen_neq} and \eqref{eq:eigen_eq} seem to take a different form.
However, if one adds a condition,
\begin{align}
  0 &= t_{\mathrm{hop}} (\Psi(l+1,l+1) - \Psi(l,l)) \nonumber \\
   &\quad + V_{\mathrm{eff}}\Psi(l,l+1), \label{eq:eigen_diff}
\end{align}
to Eq.~\eqref{eq:eigen_neq} for m = l+1, one finds that Eq.~\eqref{eq:eigen_eq} is automatically satisfied.
Note that this equation is the additional condition to derive the wavefunction for $l<m$.
Therefore, $\Psi(l,l)$ and $\Psi(l+1,l+1)$ in Eq.~\eqref{eq:eigen_diff} are generally finite.

Equation \eqref{eq:eigen_neq} is a simple difference equation, whose solution is given by the superposition of the trigonometric functions.
Using the addition theorem for trigonometric functions, we can obtain the following form of the wavefunction and the energy eigenvalue:
\begin{align}
  \Psi(l,m) =& \cos(KR+\frac{\pi}{2}r)\sin(kr+\varphi), \label{eq:exact_wave_function} \\
  E' =& 4t_{\mathrm{hop}}\sin(\frac{K}{2})\cos(k), \label{eq:exact_energy}
\end{align}
where $R = \frac{l+m}{2}$ is the center of mass coordinate, $r = m-l$ is the relative coordinate, $K$ is the center of mass momentum, and $k$ is the relative momentum.
Substituting Eq.~\eqref{eq:exact_wave_function} into Eq.~\eqref{eq:eigen_diff}, we obtain the following relation for the phase shift $\varphi$ and the momentum $k$ and $K$:
\begin{align}
  e^{2i\varphi} =& \frac{V_{\mathrm{eff}}e^{-ik}+2t_{\mathrm{hop}}\sin(\frac{K}{2})}{V_{\mathrm{eff}}e^{ik}+2t_{\mathrm{hop}}\sin(\frac{K}{2})}. \label{eq:phase_shift} 
\end{align}

The quantization conditions of the center of mass momentum $K$ and the relative momentum $k$ are given by the periodic boundary condition and the particle-hole symmetry condition as
\begin{align}
  \Psi(l,m) &= \Psi(l+L,m+L) \nonumber \\
  \implies  K &= \frac{2\pi\Lambda}{L}, \label{eq:quantization_K} \\
  \Psi(l,m) &= (-1)^{l+m} \Psi(m,l+L) \nonumber \\
  \implies  k &= \frac{2\pi \lambda + \pi - 2\varphi}{L} - \frac{\pi}{L}(\Lambda+\frac{L}{2}), \label{eq:quantization_k}
\end{align}
with $\Lambda, \lambda = -\frac{L}{2},\dots,\frac{L}{2}-1$.
We note that $K$ and $k$ are not independent of each other, since the Hamiltonian cannot be separated into the central motion part and the relative-motion part. 

Now, we are interested in the ground state, which is given by $K= -\pi$.
However, $k$ and $\varphi$ need to be solved self-consistently from Eqs.~\eqref{eq:phase_shift} and \eqref{eq:quantization_k}.
There are two types of solutions for $k$: one is the real-value solution corresponding to the scattering state, and the other is the imaginary-value solution corresponding to the bound state.

First, we consider the real value solution of $k$.
Substituting $K=-\pi$ into Eqs.~\eqref{eq:exact_wave_function} and \eqref{eq:exact_energy} gives Eqs.~\eqref{eq:dh_scattering_wf} and \eqref{eq:dh_scattering_energy} in the main text.
When $V_{\mathrm{eff}} \ll 2t_{\mathrm{hop}}$, the phase shift $\varphi$ almost vanishes, and the relative momentum $k$ is given by $k \approxeq \frac{\pi}{L}$ for the ground state.
When $V_{\mathrm{eff}} \lesssim 2t_{\mathrm{hop}}$, the phase shift $\varphi$ approaches $\frac{\pi}{2}$, and the relative momentum $k$ approaches zero.

Second, we consider the imaginary-value solution of $k = -i\kappa$.
We consider $\kappa>0$ in the following (the case of $\kappa<0$ gives the same energy and wavefunction).
From Eqs.~\eqref{eq:quantization_k} and \eqref{eq:phase_shift}, we obtain 
\begin{align}
  e^{2i\varphi} = e^{-L\kappa} = \frac{V_{\mathrm{eff}}e^{-\kappa} - 2t_{\mathrm{hop}}}{V_{\mathrm{eff}}e^{\kappa} - 2t_{\mathrm{hop}}}.
\end{align}
In the limit of $L\rightarrow \infty$, we arrive at $\kappa = \ln(\frac{V_{\mathrm{eff}}}{2t_{\mathrm{hop}}})$.
Substituting $K=-\pi$ and $k=-i\kappa$ into Eqs.~\eqref{eq:exact_wave_function} and \eqref{eq:exact_energy} gives Eqs.~\eqref{eq:dh_bound_wf} and \eqref{eq:dh_bound_energy} in the main text.

\section{Technical details of the DMRG calculation} \label{sec:Technical_Detail_DMRG}

In this Appendix, we explain the technical details of the DMRG calculation of the spinon spectrum.
For the spin system, the ground state can be expanded in terms of the local spin basis $\ket{\phi_m} = \ket{\uparrow \cdots \downarrow}$ as $\ket{s_0} = \sum_m c_m \ket{\phi_m}$.

Applying the spin removing and adding operators to the local spin basis, we obtain 
\begin{align}
  \hat{Z}_{j,\uparrow}\ket{\phi_m} =& \hat{Z}_{j,\uparrow}\lvert \underbrace{\uparrow \cdots \uparrow}_{j-1} \uparrow \underbrace{\uparrow \cdots \downarrow}_{L-2 - j} \rangle \nonumber \\
  =& (-1)^{j-1}\lvert \underbrace{\uparrow \cdots \uparrow}_{j-1} \underbrace{\uparrow \cdots \downarrow}_{L-2 - j} \rangle, \\
  \hat{Z}^{\dagger}_{j,\uparrow}\ket{\phi_m} =& \hat{Z}^{\dagger}_{j,\uparrow} \lvert \underbrace{\uparrow \cdots \uparrow}_{j-1} \underbrace{\uparrow \cdots \downarrow}_{L-1-j} \rangle \nonumber \\
  =& (-1)^{j-1}\lvert \underbrace{\uparrow \cdots \uparrow}_{j-1} \uparrow \underbrace{\uparrow \cdots \downarrow}_{L-1-j} \rangle,
\end{align}
where the factor $(-1)^{j-1}$ comes from the fermionic nature of the spinon operator.
The $(L-2)$-site spin chain changes to an $(L-3)$- (($L-1$)-)site spin chain through the spin-removal (addition) process. 
To implement this operation in DMRG, we adopt the following procedure.

First, we calculate the ground state $\ket{s_0}$ of the $L-2$ site spin chain using the DMRG method.
The tensor-network representation of the ground state MPS is given by
\begin{align}
  \ket{s_0} &=\sum_{\{\sigma\}, \{\alpha\}} A^{\sigma_1}_{\alpha_1} A^{\sigma_2}_{\alpha_1,\alpha_2} \cdots A^{\sigma_j}_{\alpha_{j-1},\alpha_j} \cdots A^{\sigma_{L-2}}_{\alpha_{L-3}} \nonumber \\
  &\quad\times\ket{\sigma_1 \sigma_2 \cdots \sigma_j \cdots \sigma_{L-2}},
\end{align}
where $\sigma_j = \uparrow, \downarrow$ is the local spin basis at site $j$, and $\alpha_j$ is the bond index.
Applying the spin-removal operator $\hat{Z}_{j,\uparrow}$, we extract the wavefunction component with $\sigma_j = \uparrow$ as
\begin{align}
  \hat{Z}_{j,\uparrow}\ket{s_0} &=(-1)^{j-1} \sum_{\{\sigma\}', \{\alpha\}} A^{\sigma_1}_{\alpha_1} A^{\sigma_2}_{\alpha_1,\alpha_2} \cdots A^{\sigma_{j-1}}_{\alpha_{j-2},\alpha_{j-1}} \nonumber \\
  &\quad\times A^{\uparrow}_{\alpha_{j-1},\alpha_j}A^{\sigma_{j+1}}_{\alpha_{j},\alpha_{j+1}} \cdots A^{\sigma_{L-2}}_{\alpha_{L-3}} \nonumber \\
  &\quad\times\ket{\sigma_1 \sigma_2 \cdots \sigma_{j-1} \sigma_{j+1} \cdots \sigma_{L-2}},
\end{align}
where $\{\sigma\}'$ means the summation over all the local spin bases except for the one at site $j$.
After relabeling the site index as $(j+1, \cdots, L-2) \rightarrow (j, \cdots, L-3)$, we obtain the ($L-3$)-site spin chain with $\hat{Z}_{j,\uparrow}\ket{s_0}$.
Similarly, applying the spin addition operator $\hat{Z}^{\dagger}_{j,\uparrow}$, we can construct the product state of $\ket{\uparrow_j} \otimes \ket{s_0}$ as
\begin{align}
  \hat{Z}^{\dagger}_{j,\uparrow}\ket{s_0} &= (-1)^{j-1} \sum_{\{\sigma\}, \{\alpha\},\beta} A^{\sigma_1}_{\alpha_1} A^{\sigma_2}_{\alpha_1,\alpha_2} \cdots A^{\sigma_{j-1}}_{\alpha_{j-2},\beta} \nonumber \\
  &\quad\times B^{\uparrow_j}_{\beta,\beta'} A^{\sigma_{j}}_{\beta',\alpha_{j}} \cdots A^{\sigma_{L-2}}_{\alpha_{L-3}} \nonumber \\
  &\quad\times\ket{\sigma_1 \sigma_2 \cdots \sigma_{j-1} \uparrow_j \sigma_{j} \cdots \sigma_{L-2}},
\end{align}
where we introduce the dummy tensor $B^{\uparrow_j}_{\beta,\beta'} = \delta_{\beta,\beta'}$ at site $j$.
After relabeling the site index as $(j, \cdots, L-2) \rightarrow (j+1, \cdots, L-1)$ and replacing $B^{\uparrow_j}_{\beta,\beta'} \rightarrow A^{\uparrow_j}_{\sigma_{j-1},\sigma_j}$, we obtain the ($L-1$)-site spin chain with $\hat{Z}^{\dagger}_{j,\uparrow}\ket{s_0}$.

Finally, we perform the Fourier transformation to calculate $N_{<}(k_s)$, $N_{>}(k_s)$, $E_{<}(k_s)$, and $E_{>}(k_s)$ as
\begin{align}
  N_{<}(k_s)=&\sum_{j} e^{ik_sj} w(j)\bra{s_0} \hat{Z}^\dagger_{j+c,\sigma} \hat{Z}_{c,\sigma} \ket{s_0}, \\
  N_{>}(k_s)=&\sum_{j} e^{ik_sj} w(j)\bra{s_0} \hat{Z}_{j+c,\sigma} \hat{Z}^\dagger_{c,\sigma} \ket{s_0}, \\
  E_{<}(k_s)=&E^{\mathrm{s}}_0 - \frac{\sum_{j} e^{ik_sj} w(j)\bra{s_0} \hat{Z}^\dagger_{j+c,\sigma} \hat{H}_{\mathrm{s}} \hat{Z}_{c,\sigma} \ket{s_0}}{N_{<}(k_s)}, \\
  E_{>}(k_s)=&\frac{\sum_{j} e^{ik_sj} w(j)\bra{s_0} \hat{Z}_{j+c,\sigma} \hat{H}_{\mathrm{s}} \hat{Z}^\dagger_{c,\sigma} \ket{s_0}}{N_{>}(k_s)} - E^{\mathrm{s}}_0,
\end{align}
where $c=\frac{L}{2}+1$ is the center-site index, and $w(j)$ is the Gaussian window function defined as $w(j) = e^{-\alpha j^2}$.
The Gaussian window function is introduced to reduce the finite-size effect originating from the open boundary condition.
Without the window function, $N_{<}(k_s)$ and $N_{>}(k_s)$ show Gibbs oscillations due to the sharp cutoff around $k_s=\pm \frac{\pi}{2}$.

\section{Mean-field analysis of the spinon spectrum} \label{sec:Mean_field_spinon}

In this Appendix, we present the spinon spectrum evaluated with the mean-field theory and discuss the difference from the DMRG analysis.
The details of the mean-field description of spinons are explained in Ref~\cite{Wen_MeanField_2007}.
The Heisenberg model with the slave-particle representation $\hat{s} = \frac{1}{2}\sum_{\alpha, \beta} \hat{Z}^{\dagger}_{\alpha}\sigma_{\alpha,\beta}\hat{Z}_{\beta}$ is given by
\begin{align}
  \hat{H}_{\mathrm{s}} = -\frac{J}{2} \sum_{l, \alpha,\beta} \hat{Z}^{\dagger}_{l,\alpha} \hat{Z}_{l+1,\alpha} \hat{Z}^{\dagger}_{l+1,\beta} \hat{Z}_{l,\beta} - \frac{J}{4} \sum_{l} \hat{n}_{l}^{Z}\hat{n}_{l+1}^{Z},
\end{align}
where $\hat{n}^{Z}_l = \sum_{\sigma} \hat{Z}^{\dagger}_{l,\sigma} \hat{Z}_{l,\sigma}$ is the spinon number operator.
Due to the constraint, $\hat{n}^{Z}_l = 1$, the last term gives a constant energy shift, $-\frac{J}{4}L$.
We perform the mean-field approximation by introducing a mean-field parameter $\chi = \ev{\hat{Z}^{\dagger}_{l,\alpha} \hat{Z}_{l+1,\alpha}}$.
By taking a real value for $\chi$ and minimizing the mean-field energy $\left< \hat{H}_{\mathrm{s}} \right>$, we obtain $\chi = \frac{1}{\pi}$.
Then, we can write down the ground state wavefunction and the mean-field Hamiltonian as
\begin{align}
  \ket{s_0} =& \prod_{|k| \leq k_F} \hat{Z}^{\dagger}_{k_s,\uparrow} \hat{Z}^{\dagger}_{k_s,\downarrow} \ket{0_s}, \\
  \hat{H}_{\mathrm{s}} =& -\sum_{k_s,\sigma}\frac{2J}{\pi}\cos(k_s)\hat{Z}^{\dagger}_{k_s,\sigma} \hat{Z}_{k_s,\sigma} + \epsilon L, \label{eq:Spinon_dispersion_mean_field}
\end{align}
where $k_F = \pi/2$ is the Fermi momentum and $\epsilon=\left(\frac{2}{\pi^2} - \frac{1}{4}\right)J$ is the energy constant per site.
Using the mean-field solution, we arrive at the spinon slave-particle spectrum,
\begin{align}
  A^{<}_{Z,\sigma}(k_s,\omega_s) &= \theta(k_F - \abs{k_s})\delta(\omega_s + \frac{2J}{\pi}\cos(k_s)+\epsilon), \\
  A^{>}_{Z,\sigma}(k_s,\omega_s) &= \theta(\abs{k_s} - k_F)\delta(\omega_s + \frac{2J}{\pi}\cos(k_s)+\epsilon).
\end{align}

\begin{figure}[t]
  \centering
  \includegraphics[width=0.45\textwidth]{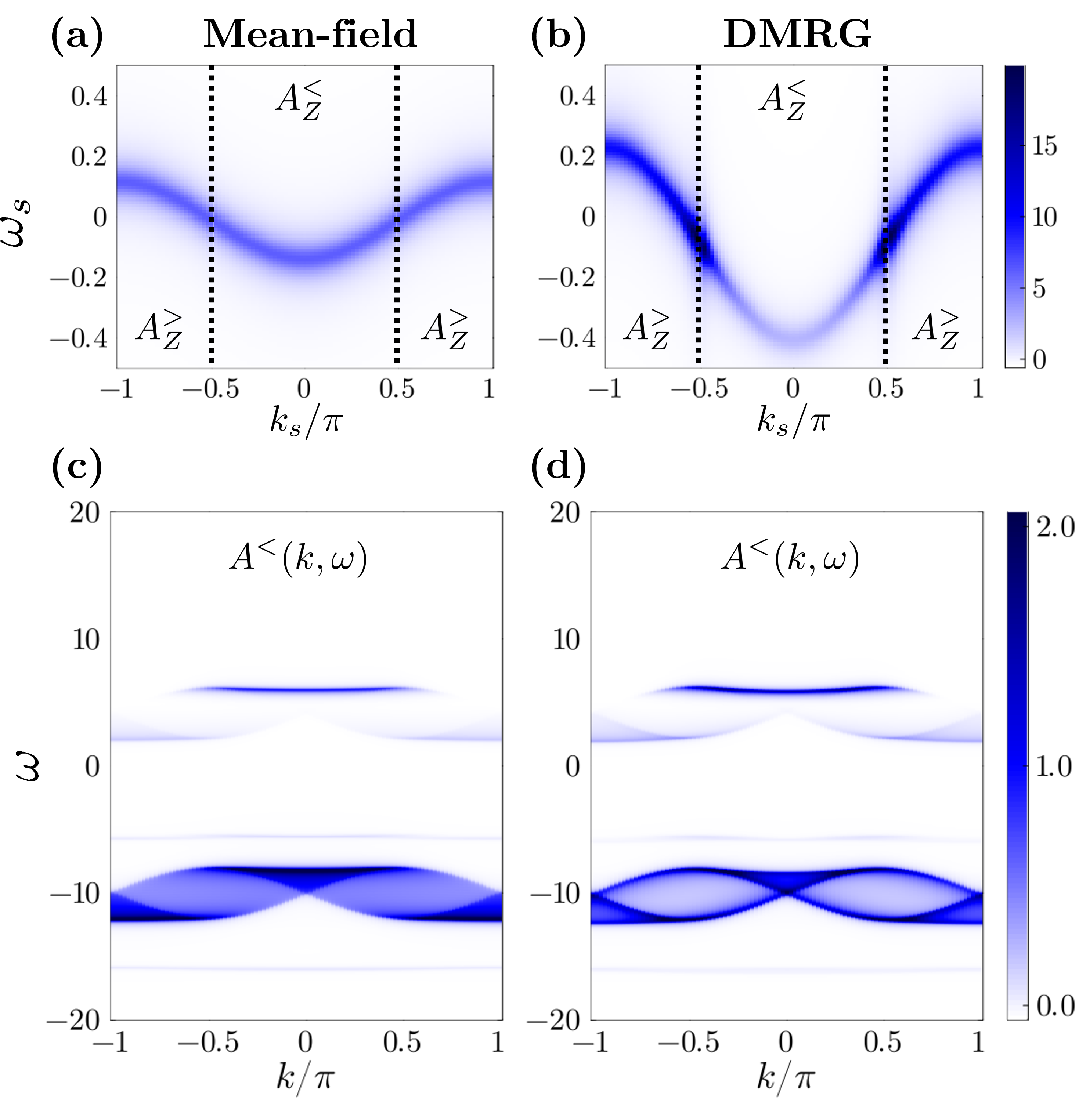}
  \caption{
    (a), (b) Spinon slave-particle spectra $A^{<}_{Z}(k_s,\omega_s)$ and $A^{>}_{Z}(k_s,\omega_s)$ calculated by (a) the mean-field approximation and (b) the DMRG method, respectively.
    (c), (d) Photoemission spectra $A^{<}(k,\omega)$ for $V=5$ calculated by (c) the mean-field approximation and (d) the DMRG method, respectively.
    For the sake of visibility, $A^{<}_{D \to S}(k,\omega)$ is multiplied by 10.
    Here we use $U=20$, $\eta=0.05$, and $L=100$.
    } \label{fig:Mean-field}
\end{figure}

To evaluate the quality of the mean-field approximation, we compare the spinon slave-particle spectrum obtained from the mean-field calculation with that from the DMRG in Figs.~\hyperref[fig:Mean-field]{13(a)} and \hyperref[fig:Mean-field]{13(b)}.
The mean-field spinon dispersion is propotional to $\cos(k_s)$, which is consistent with the DMRG result.
However, the mean-field approximation underestimates the bandwidth of the spinon dispersion by a factor of $\pi^2/4$.
Moreover, the mean-field result cannot reproduce the diverging behavior of the spectral weight at $k_s = \pm \pi/2$.

Reflecting these differences in the spinon spectrum, the photoemission spectrum calculated within the mean-field approximation shows quantitative differences from the DMRG result, as shown in Figs.~\hyperref[fig:Mean-field]{13(c)} and \hyperref[fig:Mean-field]{13(d)}.
The overall structure of the photoemission spectrum is similar, but the weight at the LHB edge (holon and antiholon branch) is less pronounced than in the DMRG result.

\bibliography{reference}
\end{document}